\documentclass[10pt,journal]{IEEEtran}
\usepackage[ruled,]{algorithm2e}
\usepackage{cite}
\usepackage{amsmath,amssymb,amsfonts}
\usepackage{algorithmic}
\usepackage{graphicx}
\usepackage{textcomp}
\usepackage{xcolor}
\usepackage{subfigure}
\usepackage{verbatim}
\usepackage{booktabs}
\usepackage{tabularx}

\usepackage[normalem]{ulem}
\definecolor{awesome}{rgb}{1.0, 0.13, 0.32}
\newcommand{\tabincell}[2]{\begin{tabular}{@{}#1@{}}#2\end{tabular}}

\def\BibTeX{{\rm B\kern-.05em{\sc i\kern-.025em b}\kern-.08em
    T\kern-.1667em\lower.7ex\hbox{E}\kern-.125emX}}

\begin{document}
\title{DeepCC: Bridging the Gap Between Congestion Control and Applications via Multi-Objective Optimization}

\author{Lei Zhang, Yong Cui,~\IEEEmembership{Member,~IEEE,}
        Mowei Wang, Kewei Zhu, Yibo Zhu,~\IEEEmembership{Member,~IEEE,} and Yong Jiang,~\IEEEmembership{Member,~IEEE,}
        % <-this % stops a space
% \IEEEcompsocitemizethanks{\IEEEcompsocthanksitem M. Shell was with the Department
% of Electrical and Computer Engineering, Georgia Institute of Technology, Atlanta,
% GA, 30332.\protect\\
% % note need leading \protect in front of \\ to get a newline within \thanks as
% % \\ is fragile and will error, could use \hfil\break instead.
% E-mail: see http://www.michaelshell.org/contact.html
% \IEEEcompsocthanksitem J. Doe and J. Doe are with Anonymous University.}% <-this % stops an unwanted space
% \thanks{Manuscript received April 19, 2005; revised August 26, 2015.}}
\IEEEcompsocitemizethanks{\IEEEcompsocthanksitem The work was supported in part by the NSFC under Grant 61872211, and in part by the National Key R\&D Program of China (No. 2018YFB1800303). \emph{(Corresponding author: Yong Cui)}\protect\\
\IEEEcompsocthanksitem L. Zhang, Y. Cui, M. Wang, K. Zhu, and Y. Jiang are with the Department of Computer Science and Technology, Tsinghua University, Beijing 10084, China (e-mail: leizhang16@mails.tsinghua.edu.cn; cuiyong@tsinghua.edu.cn; wang.mowei@outlook.com;zkw18@mails.tsinghua.edu.cn; jiangy@sz.tsinghua.edu.cn ).
\IEEEcompsocthanksitem Y. Zhu is with ByteDance Co., Ltd., Beijing 10080, China (e-mail: zhuyibo@bytedance.com).}% <-this % stops an unwanted space
% \thanks{Manuscript received XXX X, XXXX; revised XXX X, XXXX.}
}


\markboth{IEEE/ACM Transaction on Networking,~Vol.~X, No.~X, September~2020}%
{Shell \MakeLowercase{\textit{et al.}}: Bare Demo of IEEEtran.cls for Computer Society Journals}



\IEEEtitleabstractindextext{%
\begin{abstract}
  The increasingly complicated and diverse applications have distinct network performance demands, e.g., some desire high throughput while others require low latency. Traditional congestion controls (CC) have no perception of these demands. Consequently, literatures have explored the objective-specific algorithms, which are based on {\em either} offline training or online learning, to adapt to certain application demands. However, once generated, such algorithms are tailored to a specific performance objective function. Newly emerged performance demands in a changeable network environment require either expensive retraining (in the case of offline training), or manually redesigning a new objective function (in the case of online learning).
  To address this problem, we propose a novel architecture, DeepCC. It generates a CC agent that is generically applicable to a wide range of application requirements and network conditions.
  The key idea of DeepCC is to leverage both offline deep reinforcement learning and online fine-tuning. In the offline phase, instead of training towards a specific objective function, DeepCC trains its deep neural network model using
  multi-objective optimization. With the trained model, DeepCC offers near Pareto optimal policies \emph{w.r.t} different user-specified trade-offs between throughput, delay, and loss rate without any redesigning or retraining. In addition, a quick online fine-tuning phase further helps DeepCC achieve the application-specific demands under dynamic network conditions.
  The simulation and real-world experiments show that DeepCC outperforms state-of-the-art schemes in a wide range of settings. DeepCC gains a higher target completion ratio of application requirements up to 67.4\% than that of other schemes, even in an untrained environment.
\end{abstract}

}

\maketitle

\IEEEdisplaynontitleabstractindextext

% \fi
\IEEEpeerreviewmaketitle

\section{Introduction}
	
	\IEEEPARstart{T}{he} emerging applications in modern networks have very different performance requirements. Delay-sensitive applications, such as Internet telephony or cloud gaming, require a low transmission delay as low as a few milliseconds~\cite{Miller2016QoE}. 
	These applications may not benefit from higher bandwidth. On the other hand, the video streaming or file sharing, i.e., throughput-sensitive applications, often require high bandwidth for better performance~\cite{Venkat2018Copa}. In addition, some applications may provide the different specified demands of bandwidth and delay to satisfy users' quality of experience, such as some WebRTC-based applications\cite{WebRTC-performance}.
	Therefore, the transport layer should adapt to not only volatile network conditions, but also different application demands~\cite{Winstein2013Remy}.

	For the last thirty years, traditional TCP congestion controls (CC) have been dedicated to solving how to adapt to network conditions. However, they do not work well to satisfy various performance requirements due to their unaware of application demands.
% 	due to their hardwired rules. 
	For example, Cubic~\cite{XU2005CUBIC} as the default CC algorithm in Linux kernel, uses a hardwired rule to regulate congestion windows (\emph{cwnds}) and has no perception of application requirements. Many recently proposed CC algorithms, including Copa~\cite{Venkat2018Copa} and BBR~\cite{Yeganeh2017BBR}, are designed based on their own understanding of throughput/latency trade-off and share the same limitation as Cubic.
	%The recent work BBR~\cite{Neal2017BBR} highlights some new insights into the TCP modeling by probing the maximum instantaneous bandwidth and the minimum delay. However, in practice, BBR always maximizes throughput and is often apt to over-estimate the bandwidth and sacrifices latency. Thus, BBR often performs to achieve near maximum throughput. The similar results are also reported in Pantheon~\cite{Francis2018Pantheon}.
	Some other CC algorithms~\cite{Winstein2013Stochastic,Yasir2015Adaptive, Xu2013PROTEUS, Leong2017PropRate, Leong2013Mitigating}, such as Sprout\cite{Winstein2013Stochastic} and Verus\cite{Yasir2015Adaptive}, serve for specific applications or network conditions. They fall short in terms of generalization to adapt to various performance requirements. 
% 	or changed network environment.
	
	With the diverse requirements of emerging applications, the learning-based CC becomes a research hotspot recently. These learning methods are well-suited to learning control policies without relying on inaccurate assumptions. Rather than using hardwired rules, these schemes define {\em one} objective function representing a single user-specified trade-off of requirement, and learn cwnds or sending rate by optimizing the specified function.  
	For example, Remy~\cite{Winstein2013Remy} generates a decision tree to optimize an objective function of throughput and delay by {\em offline} learning. PCC~\cite{Dong2015PCC} performs {\em online} exploring for the objective optimization of throughput and loss.
	However, they are limited to optimize only one objective function with fixed parameters while do not consider the different application-specific demands. Whenever any new application requirement emerges or network environment changes, the learning-based algorithms require carefully redesigning and retuning.
    
    % \zl{Inspired by the multi-objective learning methods\cite{Yaochu2008Pareto}\cite{MTMO}, we attempt to learn various optimization functions from diverse application requirements. Along with this thinking, there are several key design challenges to be tackled, including how to guarantee the application-specific demands, how to solve the huge space of optimization objective and poor generalization problem for offline learning. }
    
	To address this problem, we propose a novel architecture, \emph{DeepCC}, where the multi-objective congestion control for {\em various} performance goals is generated through machine learning methods. When running, it automatically adapts to application-specific demands and network conditions {\em without} redesigning or retraining efforts. DeepCC is not merely built upon off-the-shelf learning methods. Instead, 
% 	it leverages two key ideas to deal with the challenges from satisfying various application requirements. 
	DeepCC realizes this with two key ideas. 
	
	First, DeepCC does not have a single fixed objective function during the training phase. The inflexibility of Remy and PCC is due to the fact that they only optimize one objective function during offline training or online learning. In contrast, DeepCC learns about how it should react to congestion with various objective functions and in various networks. DeepCC uses deep neural networks as its core model, which is well-known to be capable of handling multi-dimensional inputs and outputs. In DeepCC, these include the performance targets of throughput, delay, and packet loss, as well as the network conditions. DeepCC provides the largest possible flexibility to runtime -- different applications can specify their own performance requirements during runtime, and even change them on demand. DeepCC will still work well without retraining.

	Second, unlike the existing learning-based CC schemes, DeepCC leverages both offline deep reinforcement learning (DRL) and online fine-tuning. Offline and online approaches have their trade-offs. For example, offline approaches allow more flexible forms of objective functions but are weaker in adapting to network conditions. However, though being more adaptive to network conditions, online approaches must use special forms of objective functions~\cite{Dong2015PCC} because they will impact the online learning speed and results. DeepCC aims to take the advantages of both approaches. It learns most of its knowledge through offline training, which is more focused on supporting flexible objective functions, while still includes an online fine-tuning phase that handles dynamic network conditions. This online fine-tuning is much more efficient than a purely online design because the results from offline training provide a good starting point and largely narrow down the search space. The main difference between DeepCC and other schemes are summarized in Table \ref{table1}.
	
	To the best of our knowledge, DeepCC is the first congestion control that can optimize the multiple objective functions. It is also the first CC to combine both offline learning and online fine-tuning. Though this approach is less seen in the networking community, it is in fact popular in the machine learning research and has achieved many state-of-the-art results~\cite{Silver2016Mastering,Devlin2018BERT}.  The main contributions are listed as follows:

	\begin{itemize}
		\item We present a novel architecture named DeepCC, which can satisfy different performance requirements without redesigning or retraining efforts. It fully leverages the power of offline and online learning techniques that improve the generalization ability for both application requirements and network conditions (\S\ref{sec:overview}).

		\item We propose a multi-objective DRL algorithm to learn Pareto optimal (or near optimal) control policies for different performance trade-offs. Our solution efficiently explores the wide optimization objective space and offline learn an optimal (or near optimal) strategy (\S\ref{design-multiobj} and \S\ref{design-offline}).

		\item We design an online tuning algorithm for various application-specific demands and network conditions. It can dynamically choose one of the learned Pareto optimal policies to adapt to the real-time network conditions under the guidance of the specific demand. It greatly facilitates DeepCC to meet the explicit performance requirements under different network environments (\S\ref{design-online}).
		
% 		We evaluate DeepCC through packet-level network emulator Mahimahi~\cite{Netravali2015Mahimahi} over network traces. To validate the performance of DeepCC, we compare it against other existing \zl{state-of-the-art} algorithms. The experiment results show that DeepCC can achieve a wider range of trade-offs between throughput, delay and packet loss than that of existing algorithms. Once being trained, DeepCC can adaptively tune objective to achieve or be close to the target value of application requirements in different network scenarios. DeepCC outperforms existing \zl{learning}-based algorithms with the improvements in coverage ratio of various application requirements ranging from {4.80\% to 68.20\%}, even on the network scenarios for which it was not trained. 
	\end{itemize}
	We implement and evaluate DeepCC in the Mahimahi~\cite{Netravali2015Mahimahi} emulator and real-world network. Compared against state-of-the-art schemes, DeepCC achieves a wide range of performance and gains a higher target completion ratio (TCR) of application requirements up to 67.4\% than that of other schemes, even in an untrained environment
% 	The results over untrained network conditions are summarized in Table \ref{table1}. 
% 		\zl{DeepCC outperforms other algorithms to achieve a wide range of performance and improve the target completion ratio (TCR) of various application requirements}
    %  It achieves a wide range of performance and high target completion ratio  to support various application requirements
		(\S\ref{sec:impl} and \S\ref{sec:evaluation}).
% 	The DeepCC gains a higher target completion ratio of application requirements up to 67.4\% than that of other schemes, even in an untrained environment.

	\begin{table}[t]
	    \renewcommand\arraystretch{1.3}
        \centering  
	    \caption{Differences between DeepCC and other schemes}  
	    % \vspace{-0.07in}
        \label{table1} 
	   % \small
	    \begin{tabular}{p{1.35cm}<{\centering}|p{1.4cm}<{\centering}|p{1.65cm}<{\centering}|p{1.3cm}<{\centering}|p{1.0cm}<{\centering}}
		    \hline 
		     Solution&\tabincell{c}{Optimization\\ objective}&Offline/Online&\tabincell{c}{Diverse\\requirements}&TCR(\%)\\  
		    \hline
		    Cubic\cite{XU2005CUBIC} & / & / & $\times$ & 0.0\\  
	    	\hline
		    BBR\cite{Yeganeh2017BBR} & / & / & $\times$ & 0.0\\ 
		    \hline
		    Remy\cite{Winstein2013Remy} & Single  & Offline & $\times$ & 44.20  \\
		    \hline
		    PCC\cite{Dong2015PCC} & Single  & Online & $\times$ & 0.0\\ 
		    \hline
		    Vivace\cite{Dong2018Vivace} & Single  & Online & $\times$ & 0.0\\ 
		    \hline
		    Indigo\cite{Francis2018Pantheon} & Single  & Offline & $\times$ & 3.20\\ 
		    \hline
		    DeepCC & Multiple &Offline\&Online& $\surd$ & 67.40\\ 
		    \hline
	    \end{tabular}
	   % \vspace{-0.15in}
    \end{table}
% 	The remainder of this paper is organized as follows. We introduce the related work in Section II. We present our motivation in Section III. We present the system overview and detailed design in Section VI and Section V. We evaluate the performance of our solution in Section VI and conclude our work in Section VII.
	
\section{Motivation and Challenge}
    Comparing with traditional congestion control schemes, learning-based approaches can adapt to network conditions, such as Remy\cite{Winstein2013Remy}, Indigo\cite{Francis2018Pantheon}, PCC\cite{Dong2015PCC}, and Vivace\cite{Dong2018Vivace}. Nevertheless, poor generalization ability limits them to work with changing application-specific requirements and network conditions. In the following, we first summarize the limitations of the existing schemes. Then we detail the key challenges that are tackled by DeepCC's design.
    
    % Comparing with traditional congestion control schemes, objective-based approaches take one step further by providing network operators with an objective function to easily define specific optimization objectives. Nevertheless, diverse requirements and poor generalization ability limit them to work with changing applications and network conditions. In the following, we first summarize the limitations of current objective-based schemes. Then we introduce the opportunities that come along with multi-objective optimization.
    
    % We then introduce the opportunities that multi-objective optimization may bring in.
    
    % 	Unlike the traditional CC, the objective-based approaches offer a potentially flexible proxy (i.e., the objective function) to achieve performance objective. Thus we focus on the objective-based approaches for the following comparison to show their limitations in terms of generalization to satisfy various application requirements and network conditions. 
    
     \begin{figure*}
        %\vspace{-0.1in}	
        \centering
        % \vspace{-0.15in}
        \subfigure[Cellular network]{
            \label{motivation-att} 
            \includegraphics[width=0.315\linewidth]{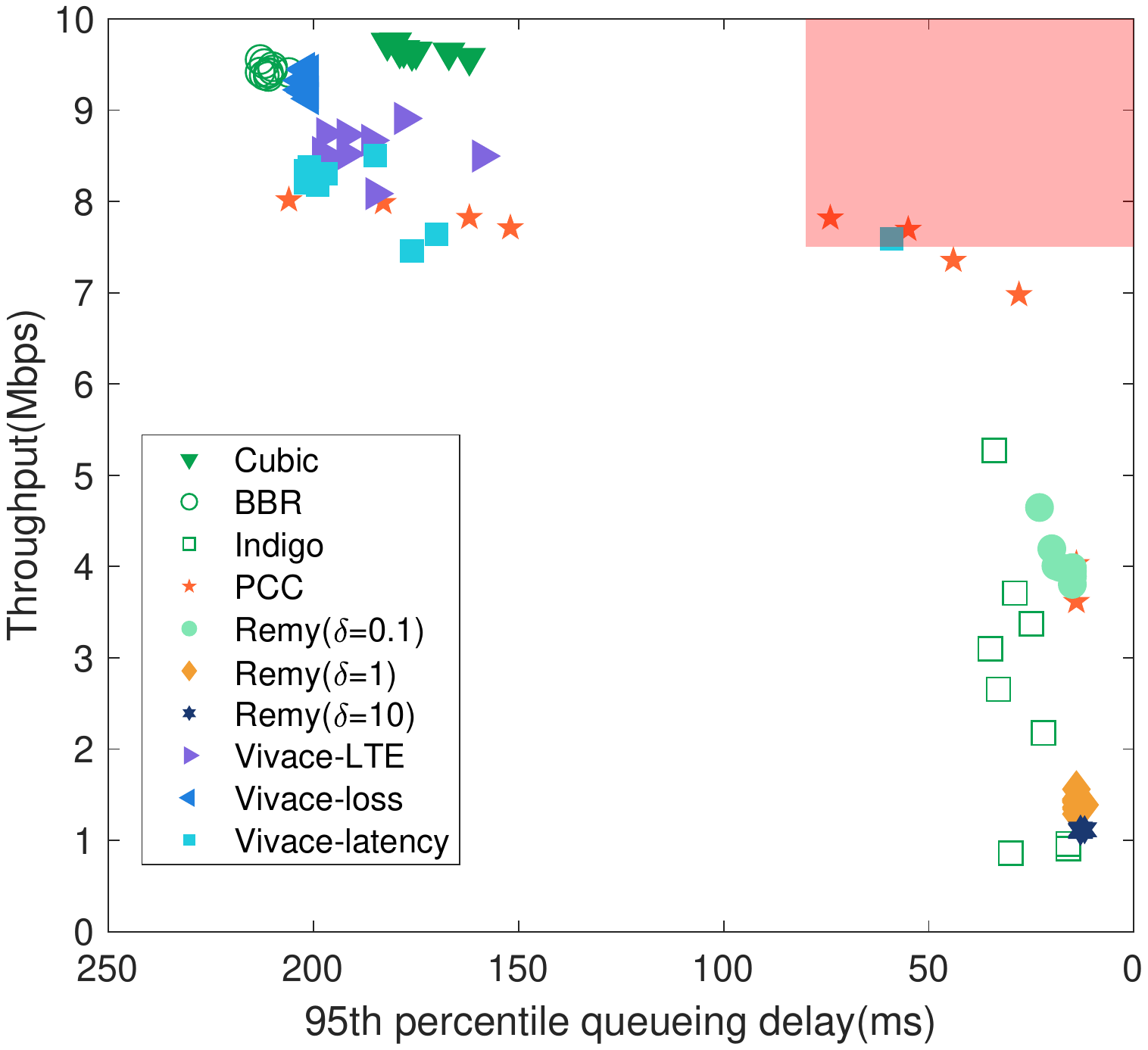}}
        \hspace{0.01in}
    	% \vspace{-0.1in}	
        \subfigure[Wi-Fi network]{
            \label{motivation-wifi} 
            \includegraphics[width=0.32\linewidth]{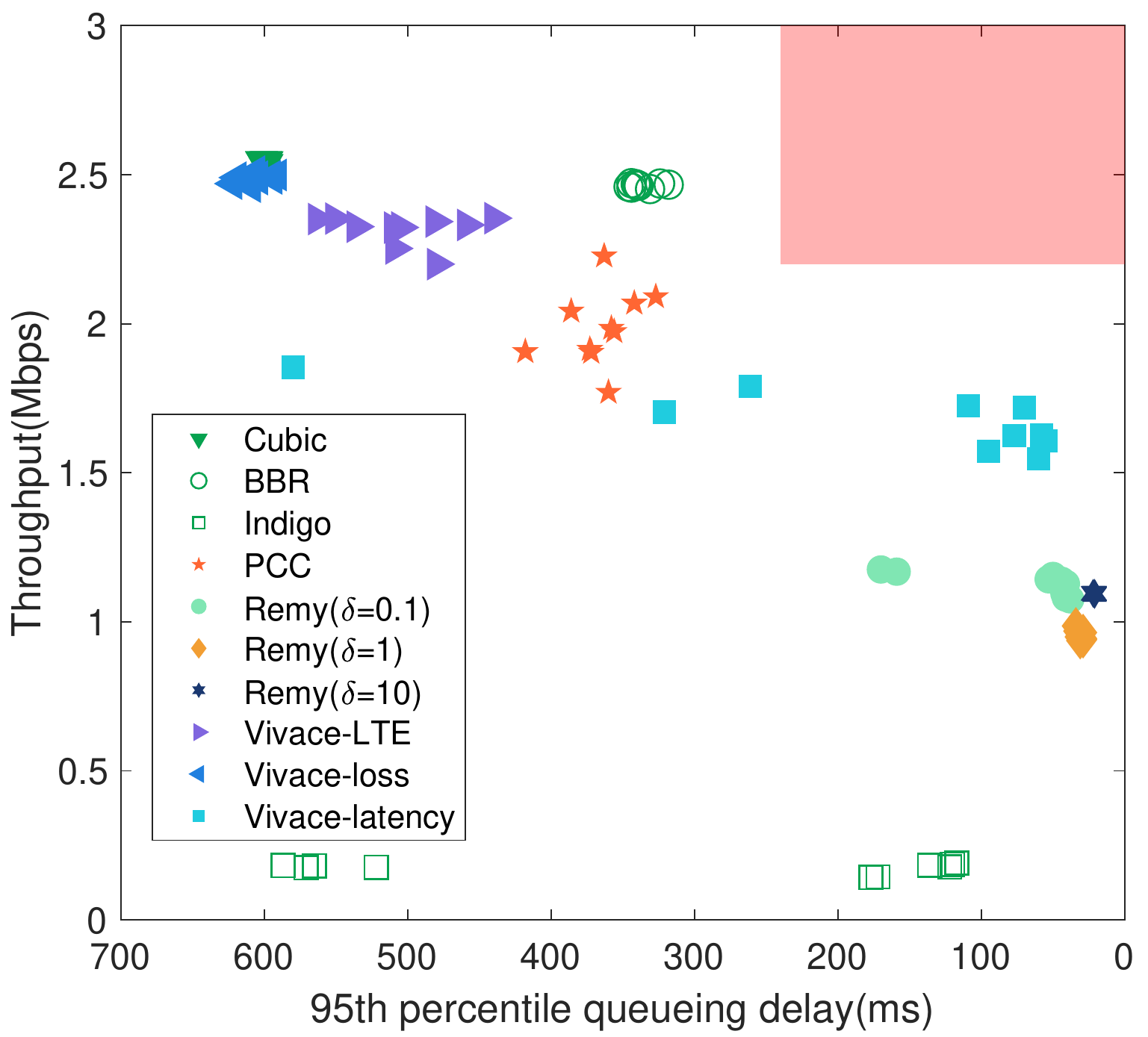}}
        % \vspace{-0.15in}	
        \subfigure[Wired network]{
            \label{motivation-12} 
            \includegraphics[width=0.32\linewidth]{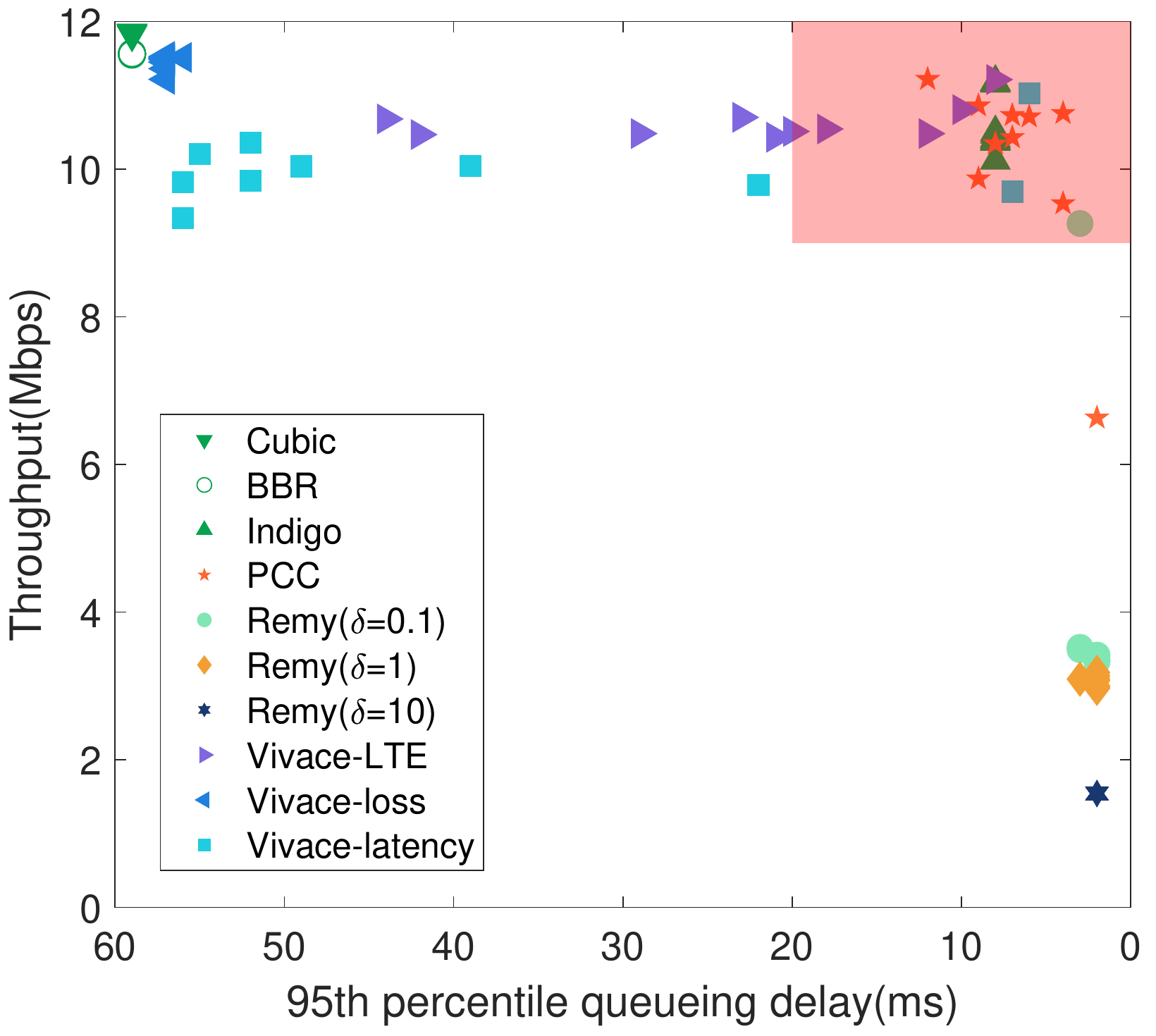}}
        \caption{Throughput vs. 95th percentile queueing delay achieved by the existing CC with their objective functions under three network links. All schemes are repeatedly tested 10 times and each test lasts for 50 seconds.}
        \label{unpredictable_performance} 	
        % \vspace{-0.15in}	
    \end{figure*} 
    \subsection{Limitations of the existing schemes}
    	\label{limitation}
    	We conduct experiments to evaluate the performance of {existing state-of-the-art} schemes\footnote{The source codes are provided by Pantheon~\cite{Francis2018Pantheon}.} in cellular and {Wi-Fi links}. We plot the distribution of throughput and 95th percentile queueing delay of them in Fig.~\ref{unpredictable_performance}. We then highlight the limitations with illustrative examples. 

    	\textbf{\emph{Limitation 1: Existing schemes can not generalize for diverse application requirements. }}
    	
    	Manual policies, such as Cubic\cite{XU2005CUBIC}, allow the senders to deliver data based on heuristics rather than performance requirements. Application-specific optimization schemes, such as Remy and PCC, often do not have general applicability for diverse performance requirements. 
    	A fine-tuned scheme for one application works poorly for other applications if performance requirements change.
    
        % \textbf{{Limited choices of objective functions. }}
    	In principle, the learning-based schemes can support application requirements by designing their objective functions. However, the existing schemes, such as Remy or Vivace, at least in their current form, only provide three choices of objective functions. As a direct consequence, they work in some isolated operating points. 
        As shown in Fig.~\ref{unpredictable_performance}, the family of Vivace can only operate on the high-throughput area (the left-top) and cannot achieve low latency, while that of Remy fails to escape from the scope of low-latency area (the right-bottom). None of them could cover the Pareto front\cite{Yaochu2008Pareto}. Although Remy and Vivace can behave on different trade-offs with their three objective-specific functions, the performance range is quite limited so that they can not satisfy diverse requirements.
        
        % \textbf{Unpredictable performance. }
    	Even if there are infinite objective functions available that could achieve any trade-offs, it is still non-trivial for the applications to use such a learning-based scheme. The main reason is that none of these schemes provide a systematic approach that could guide the CC to select the suitable objective function to meet a specific  quality of service (QoS) requirement (e.g., 8 Mbps throughput while keeping the delay within 70 ms and the loss rate within 1\%, the shaded area as shown in Fig.~\ref{motivation-att}). Moreover, the average queueing delay (calculated as the difference between the observed round-trip time, i.e., RTT, and the minimum RTT) and throughput of a single scheme with one objective function can be highly variable under a single network scenario in Fig.~\ref{motivation-att}, such as PCC or Indigo. Hence, in the view of the application providers, the existing learning-based schemes may be still far from satisfactory.
    	
    	\textbf{\emph{Limitation 2: 
    	Existing learning-based schemes cannot generalize well across a wide range of network conditions.}}
    	
    	Except the application-specific demands, congestion control needs to adapt to a wide variety of heterogeneous network conditions.
    	The existing online or offline schemes can benefit from their own properties but still face great limitations.
    	
    % 	\textbf{Limited generalization for offline schemes. }
    	Schemes relying on offline learning inherently face the generalization problem. The model or agent (i.e., the learned control rules) can only be trained with limited data that cannot cover all the network conditions and thus may overfit the dataset, at least to some extent. For example, Indigo could attain low latency within its design scope in Fig.\ref{motivation-12}. However, its performance can be degraded when the actual network conditions mismatch the training assumptions~\cite{sivaraman2014experimental} as shown in Fig.\ref{motivation-wifi} .
    % 	The performance of Indigo in Fig.~\ref{unpredictable_performance} provides another evidence for this. Indigo has been trained on the steady link by imitating the optimal sending action. When facing the network scenario where the bandwidth is highly variable like cellular link, it performs far from acceptable.
    	
    % 	\textbf{Slow convergence for online schemes.}
    	Although the online schemes perform at least acceptably when facing new network condition, the convergence time is a fundamental problem for them. We compare the convergence behavior of the above-mentioned schemes as shown in Fig.~\ref{convergence time}. It is similar to the results that Vivace claims - PCC converges slowest~\cite{Dong2018Vivace}. In general, the offline schemes are more stable than online schemes\cite{schapira2017congestion}. This is because the offline schemes can leverage prior knowledge of networks obtained during training.
    % \vspace{-0.05in}
    % \subsection{Opportunity}
    %     The objective-based congestion control schemes with fixed objective function have challenges in (1) expressing diverse application requirements in an apprehensible manner, and (2) adapting quickly to unknown network conditions. 
    %     Fortunately, these problems can be potentially solved by new techniques with the machine learning community.
    %     Recently, deep reinforcement learning (DRL) has achieved impressive performance breakthroughs in many fields~\cite{Silver2016Mastering,Mao2016Resource,li2017deep}, which has high potential in learning complex patterns from the past experiences. Meanwhile, online learning has the ability to adapt to environment quickly.
    %     Inspired by the advantages of DRL and online learning, we prospect that offline DRL with multi-objective optimization can learn multiple policies when meeting different requirements. Then by leveraging online learning, we expect that our approach could adapt to different network conditions by choosing the proper  offline-learned policy. 
    %     % which are trained to optimize different objectives}.
    %     % Whereas, online learning has the ability to quickly adapt to unseen network conditions.
%  \vspace{-0.15in}

    \subsection{Challenge}
    \label{sec:challenge}
    As many have observed, learning-based congestion control schemes have emerged in support to adapt to complicated network conditions.
    % The fine-tuned policies allow developer to trade application-specific requirements. However, it's difficult to write accurate policies without extensive domain expertise or considerable effort for diverse application demands. 
    Although these schemes deliver a satisfactory performance of a single specific requirement, they still have limitations and fail to achieve different trade-offs of performance requirements. 
    Along this direction, DeepCC is designed to tackle the following key challenges.
    
    (1) {Guaranteeing the application-specific demands. }
    It is important to guarantee the specified demands for the applications. However, achieving the specified performance is non-trivial because CC schemes achieve the performance unpredictably under complicated network conditions. Further, learning-based schemes could provide a desired trade-off for one or a class of applications by defining a fine-tuned objective function, but it is difficult to achieve the deterministic performance demands defined by applications or users. 
    
    (2) {Huge space of optimization objectives. }
    Each metric of application requirements could span on a large scale. 
    The diverse requirements with multiple dimensions that fall into different trade-offs of the performance metrics further make it harder to deal with.
    % The diverse performance requirements make the optimization space grow exponentially. 
    The huge space of optimization objectives poses a great challenge, especially for reinforcement learning, which must "explore" the action space in training to learn a good policy for each optimization objective.  
    
    \begin{figure*}[t]
    	\centerline{\includegraphics[width=0.95\linewidth]{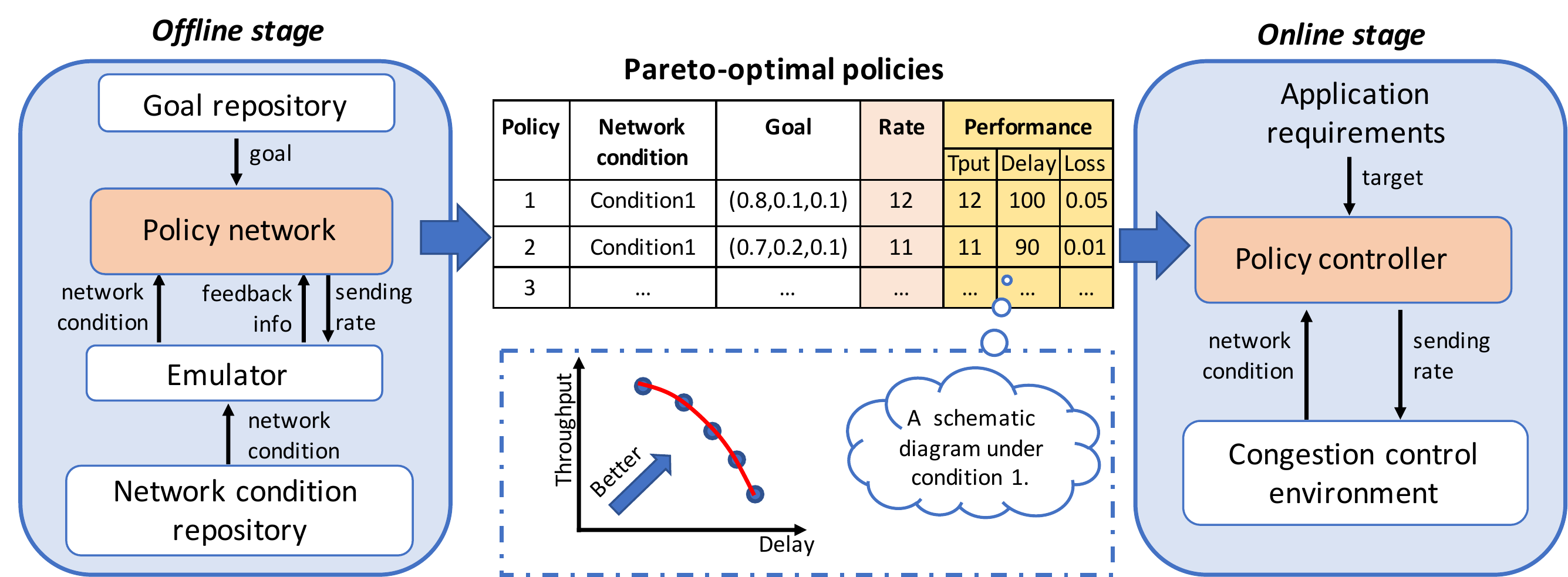}}
    	% \vspace{-0.15in}
    	\caption{\label{overview}The DeepCC architecture.}
    % 	\vspace{-0.15in}
    \end{figure*}
    
    (3) {Poor generalization for learning-based schemes. }
    Online learning has the ability to adapt to the environment quickly, whereas the convergence time is too long. 
    By contrast, the learned agent or model through offline learning, which is provided a good starting point by offline learning, can perform well in the scenarios that are similar to the training environments. But it could fail to adapt to unseen network conditions. 
    % The first challenge is that how to learn a single DRL agent to quickly meet both diverse application requirements and complicated network conditions. 

\section{Design Overview}
\label{sec:overview}
{
In this work, we seek to close the gap between the congestion control and different applications' requirements by proposing DeepCC, a multi-objective CC with various optimization \emph{goals} {that can adapt to meet different} \emph{targets} of application-specific demands (See Table \ref{Notation} for an explanation of the notation). DeepCC does not need to modify the TCP protocol and only {obtains} the QoS requirements from the upper layer applications.
DeepCC is not merely {built upon off-the-shelf reinforcement learning approaches}. Instead, it {leverages several ideas to solve the above-mentioned challenges}. }

In conventional learning-based congestion control, {the agent} can collect network states from environments and {learn to improve its output action} based on its fine-tuned optimization objective. However, {this approach is unaware of the different} application demands {so that its achieved performance can be away from the actual application requirements}. 
% However, it is insensitive to the application demands and achieves the performance unpredictably. 
To cope with the problem, we propose an {online algorithm} to fulfill the application-specific demands (e.g., 10 Mbps throughput while keeping the delay within 100 ms and the loss rate within 1\%). The {online algorithm} allows DeepCC to adaptively tune the policy according to the application demands. 

The learning-based agent can explore diverse network environments to enrich its experience and {try to meet} all metrics of application requirements, including different throughput, delay, and loss rate. 
% Intuitively, it is difficult to train an optimal or near-optimal agent from massive application requirements directly. 
{However}, the huge space of optimization objectives makes it very difficult to train an optimal or near-optimal agent {for} massive application requirements. To handle this problem, DeepCC does not directly optimize for diverse performance requirements instead it optimizes for different trade-offs of performance metrics as an intermediate objective, {using the weights to represent different trade-offs}. Specifically, DeepCC leverages multi-objective optimization to offline generate a flexible agent that is able to learn the Pareto optimal or near-optimal policies {and be tuned for different objectives after training.}

{It is well known that poor generalization is a key problem of learning-based schemes. To mitigate the generalization problem, we propose the two-stage learning architecture which leverages the benefit from both offline and online learning. In our cases, even if the same application requirements, DeepCC should adopt different policies in different network environments. In this way, DeepCC continuously adjusts the weights of the model obtained from offline training through online tuning algorithm to meet the application's target under different environments.
Fig.~\ref{overview} shows the high-level overview of DeepCC's design which contains two stages.  
}

\begin{table}[t]
	    \renewcommand\arraystretch{1.3}
        \centering 
 	    \caption{Notation}  
	    % \vspace{-0.1in}
        \label{Notation}  
	    \begin{tabular}{p{1.5cm}|p{4.8cm}|p{0.7cm}}
		    \hline  
		     Name  & Description & Symbol  \\  
		    \hline
		    State  & the last action & $s^{(1)}$  \\ 
		      &  an exponential weighted moving average of the current RTT & $s^{(2)}$ \\  
		      &  an actual sending rate & $s^{(3)}$ \\ 
		      &  the average delay in each decision interval & $s^{(4)}$ \\ 
	    	\hline
		    Action & the sending rate & $a$ \\ 
		    \hline
		    Reward & the compound of measurement and goal & $r$ \\
		    \hline
		    Goal & the relative weight of throughput & $g^{(1)}$ \\ 
		     & the relative weight of delay & $g^{(2)}$ \\ 
		     & the relative weight of packets loss rate & $g^{(3)}$ \\ 
		    \hline
		    Measurement  & the normalized throughput & $m^{(1)}$\\ 
		     & the normalized delay & $m^{(2)}$ \\ 
		     & the observed packet loss rate & $m^{(3)}$\\ 
		    \hline
		    Target  & the requirement of minimum throughput & $T^{(1)}$\\ 
		     & the requirement of maximum delay & $T^{(2)}$ \\  
		     & the requirement of maximum loss rate & $T^{(3)}$ \\ 
		    \hline
	    \end{tabular}
	   % \vspace{-0.15in}
    \end{table}

In the offline stage, DeepCC learns a set of Pareto optimal policies (i.e., the policy network with different goals as input condition in Fig.~\ref{overview}) under different network conditions.
% 	DeepCC represents the congestion control algorithm as an agent that uses deep neural networks to make decisions, referred to as the \emph{policy network}.
Considering the diverse performance requirements, we start by defining a group of weights, which expresses the different trade-offs (termed \emph{goal}) of the relative preferences for throughput, delay, and loss rate  (\S\ref{design-multiobj}). The policy network takes the goal and the network conditions (termed \emph{state}) as input and outputs the sending rate. DeepCC trains the policy network to optimize for different goals through a large number of offline experiments from the emulator (\S\ref{design-offline}). The well-learned policy network can build a good relationship between the input conditions and the sending rate over all possible performance trade-offs. 
    
In the online stage, DeepCC matches the network condition and performance requirement (termed \emph{target}) to the sending rate (termed \emph{action}) using the learned Pareto optimal policies. 
At the start of the connection establishment, applications provide their requirements of bandwidth, delay, and packet loss to the policy controller (\S\ref{design-online}). At runtime, the controller continuously detects the changes in the difference between current performance and requirements, and automatically chooses the most proper policy (represented by \emph{goal}) that can best fit the requirements. Then the best sending rate is decided according to the selected policy.

% \vspace{-0.1in}

\section{Detailed Design}
	In this section, we present the detailed design of DeepCC. We begin with describing the multi-objective function.
	Then we explain the offline training process with multi-objective optimization and the online tuning algorithm.
 
% \subsection{Multi-Objective Reward Function}
\subsection{Representing multi-objective function}
	\label{design-multiobj}
	
	Note that it is non-trivial to directly learn to achieve the various performance requirements offline, since the application requirements may fall in a large performance space especially when considering different network conditions. 
    Hence, we use the relative instead of the absolute value to express our objective function in offline learning. Further, the \emph{goal} as a relative weight of the performance metrics can become a direct ``knob" to be tuned by users or applications online to achieve the application desired performance.
	
	The multi-objective expression includes not only multi-dimensional performance metrics, i.e., throughput, delay, and loss rate, but multiple trade-offs between them. The multi-objective function is composed of {measurement} and {goal}. Among them, the \emph{measurement} indicates the current transmission performance that includes the throughput, delay, and loss rate. Table ~\ref{Notation} summarizes these symbols.
	
	\noindent \textbf{Measurement. }The measurement $m$ is an $n$-dimensional vector as $m=(m^{(1)}, m^{(2)}, \cdots ,m^{(n)})$.
	Considering the performance metric after the action taken in congestion control, we set $n=3$ and the $m_{t}$ at time step $t$ as:
	\begin{equation} 
	m_{t}=(\frac{throughput_{t}}{throughput_{max}}, \frac{delay_{t}}{delay_{min}}, loss\ rate_{t})
	\label{eq4}
	\end{equation}
	At time step $t$, the $throughput_{t}$ is the instantaneous observed total throughput of the sender and the $throughput_{max}$ is the maximum value among all the history throughput; $delay_{min}$ is the minimum delay of the current connection; $delay_{t}$ is the 95th percentile delay; $loss\ rate_{t}$ is the observed loss rate. 

	\noindent\textbf{Goal. } The goal $g_{t}$ is also an $n$-dimensional vector as:
	\begin{equation}
		 g_{t}=(g_{t}^{(1)}, g_{t}^{(2)}, \cdots ,g_{t}^{(n)})    \qquad s.t. \sum_{i=1}^{n}g_{t}^{(i)}=1
	\vspace{-0.07in}
	\end{equation}
	where $g_{t}^{(i)}$ is the relative weights of the corresponding performance metric at time step $t$. And the sum of all $g_{t}^{(i)}$ equals one. In CC problem, the \emph{goal} represents the different trade-offs between throughput, delay and loss rate, i.e., $n=3$. The larger $g_{t}^{(1)}$ signifies that higher throughput is preferable. The larger $g_{t}^{(2)}$ and $g_{t}^{(3)}$ indicate that the lower delay and loss rate are preferable respectively. 
	
	\noindent\textbf{Reward. }
    Similar to the reward function of DRL, Remy and PCC use the objective function or the utility function to evaluate the transmission performance and take them as the feedback that helps decision making. However, they only use two of the three performance metrics. Specifically, their objective function includes multiple dimensions rather than the different trade-offs between the performance metrics.

	In our approach, the multi-objective function, i.e., \emph{reward}, is set to reflect the desired performance of throughput, delay, and loss rate that we wish to optimize under different relative weights. So we set the reward $r_{t}$ as the compound of measurement and goal at each time step $t$. For congestion control, we set the performance of throughput as an award, while the performance of delay and loss rate as a penalty. Therefore, when computing a reward, the 2nd-dimension and 3rd-dimension of measurement use the \emph{negative} value of them.
% 	\vspace{-0.02in}
% 	\begin{small}
	\begin{equation}
	r_{t}= g_{t}^{(1)}m_{t}^{(1)}-g_{t}^{(2)}m_{t}^{(2)}-g_{t}^{(3)}m_{t}^{(3)}/threshold \label{eq5} 
	\vspace{-0.03in}
	\end{equation}
% 	\end{small}
	
    \noindent{The $threshold$ represents the tolerance of packet loss. }By explicitly introducing multi-objective reward, the agent learns multiple objectives with different goals.
    Moreover, the decision-making does not depend on the intermediate reward in one step, but takes the expected cumulative reward $J_{mul} = \mathbb{E}[\sum_{t=0}^{N}\gamma^{t}r_{t}]$ as the objective, where $\gamma \in (0,1)$ is a discount factor and $N$ is the total steps.

\subsection{Offline learning with multi-objective DRL}
    % \subsubsection{State and Action}
	\label{design-offline}
	\begin{figure}[t] 
		% \vspace{-0.1in}
			\centerline{\includegraphics[width=0.99\linewidth]{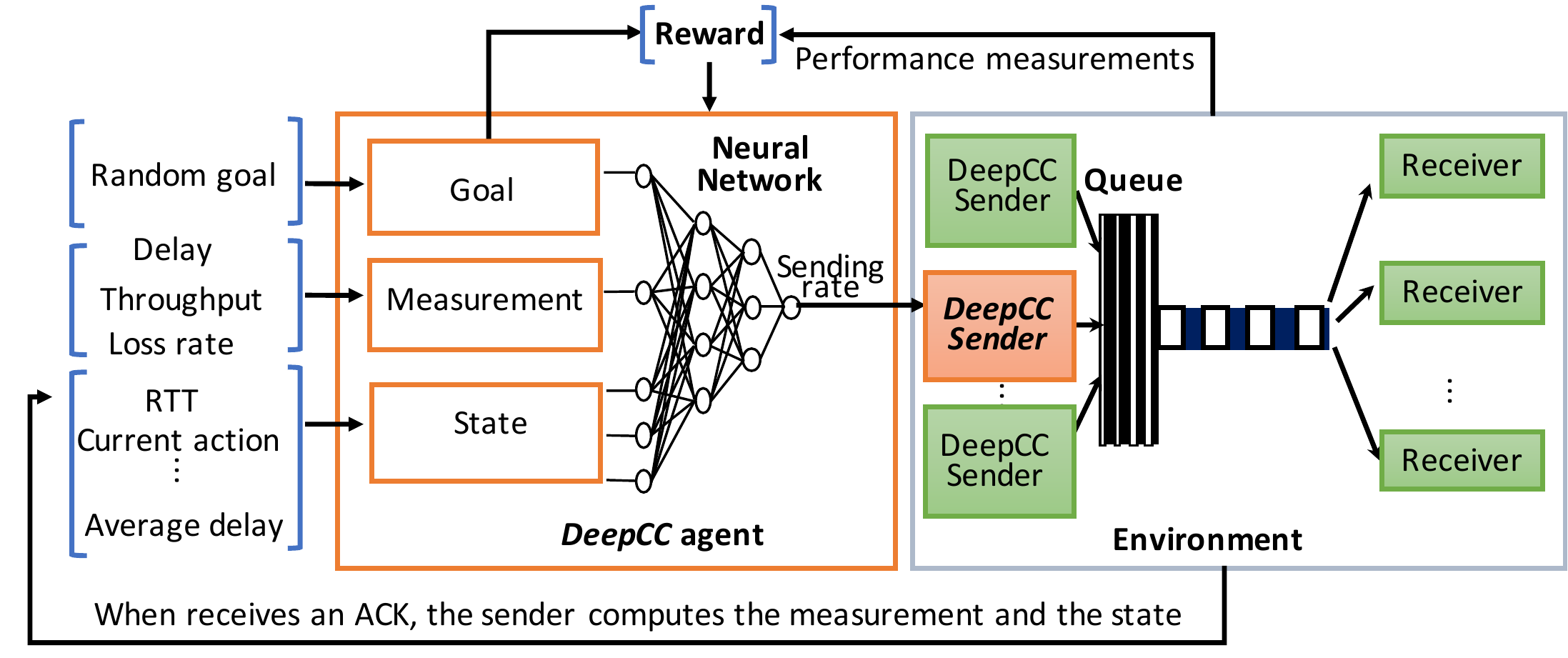}}
			% \vspace{-0.1in}
			\caption{\label{offline-architecture} Offline learning with multi-objective optimization.}
% 			\vspace{-0.15in}
	\end{figure}
	
% 	\noindent\textbf{Offline stage:}
    We consider a DeepCC agent generated by the multi-objective DRL through offline learning, which can potentially deal with the multiple objectives, the continuous decision space, and the adaptation problem by leveraging the great power of deep neural networks (NNs). 
    Unfortunately, the basic Deterministic Policy Gradient (DDPG) algorithm ~\cite{Timothy2016DDPG}, an advanced DRL algorithm that deals with the continuous action space, could not directly support for DeepCC with multi-objective optimization due to its limited expressiveness of the fixed scalar reward. Recently, some novel multi-objective reinforcement learning algorithms have been proposed, such as DFP~\cite{Alexey2017DFP} for the game Doom and UVFAs~\cite{Tom2015UVFAs} for Atari games. However, both DFP and UVFAs cannot directly be applied to continuous control problems. 
	
	Hence, we design a multi-objective DDPG algorithm based on basic DDPG architecture. Specifically, we use a multi-objective function as the reward function of DeepCC (provided by \S\ref{design-multiobj}), expressed as the combination of the {goal} and performance metric {measurement}. In the offline training process, we set a \emph{random goal} for each training episode to achieve a wide range of trade-offs between the performance metrics. 

	Unlike traditional CC schemes that use hardwired rules to regulate cwnds or sending rate, DeepCC agent learns the flexible policy of sending rate directly from the agent-environment interactions. As shown in Fig.~\ref{offline-architecture}, 
% 	the agent observes the \emph{state} from the network environment, takes an \emph{action}, and receives a \emph{reward}. The \emph{state} is a set of observations including the current RTT, the average delay, the actual sending rate, and the last action.
% 	The \emph{action} in DeepCC corresponds to the sending rate, which is controlled by a {policy} neural network.
	the input of agent includes not only network \emph{state} but also \emph{measurement} and \emph{goal}. Then the agent derives the proper \emph{action}, i.e., sending rate, when receiving an ACK. 
	Moreover, the multi-objective function of the agent uses the compounded information of measurement and goal which is used to train and improve the neural network model.
	The ultimate aim of the learning algorithm is to maximize the expected cumulative discounted reward.
	Leveraging the power of DNNs, the agent can learn near-optimal control policies that mapping state, measurement, and goal to the action for each optimization objective.
	
	Formulating congestion control problem as a DRL task in DeepCC requires specifying the {state} and {action}.

	\noindent\textbf{State space. }
	\label{design-state}
	When the sender receives an ACK, the agent observes the current RTT and computes the history statistics. We narrow our attention to some statistics and observations as the state $s_{t}=(s_{t}^{(1)},s_{t}^{(2)},s_{t}^{(3)},s_{t}^{(4)})$ that may facilitate the CC decisions. The detailed descriptions of $s_{t}^{(i)}$ are in Table \ref{Notation}.
% 	i.e., $s_{t}^{(1)}$ is the last action; $s_{t}^{(2)}$ is an exponential weighted moving average of the current RTT; $s_{t}^{(3)}$ is an actual sending rate which is defined as the number of packets sent since the last ACK has been received divides by the current RTT; $s_{t}^{(4)}$ is the average delay in each decision interval.

	\noindent\textbf{Action space. } 
	% Traditional cwnd-based CC mechanisms (e.g., Cubic~\cite{XU2005CUBIC}) bring about large latency due to burst transmission. The rate-oriented CC can directly regulate the packets of senders to alleviate the bufferbloat problem.
	% Therefore, 
	We choose the sending rate, a continuous variable, as the action of the agent. After receives $s_{t}$, the agent takes the action $a_{t}$, i.e., the sending rate. The action is selected by a policy $\mu(s_{t})$ which is defined as a deterministic action $a_{t}\in [0,bound]$, where $bound$ is the upper limitation of the sending rate. 
    % \zl{to avoid the potentially unpredictable behaviors}. 
	We use NNs to represent the policy with a manageable number of adjustable parameters $\theta^{\mu}$.
% 	\doubt{I thought "avoid the potentially unpredictable behaviors" seems convey little information here.}
    % \vspace{-0.05in}
    
\begin{figure}[t] 
			% \vspace{-0.1in}
			\centerline{\includegraphics[width=0.99\linewidth]{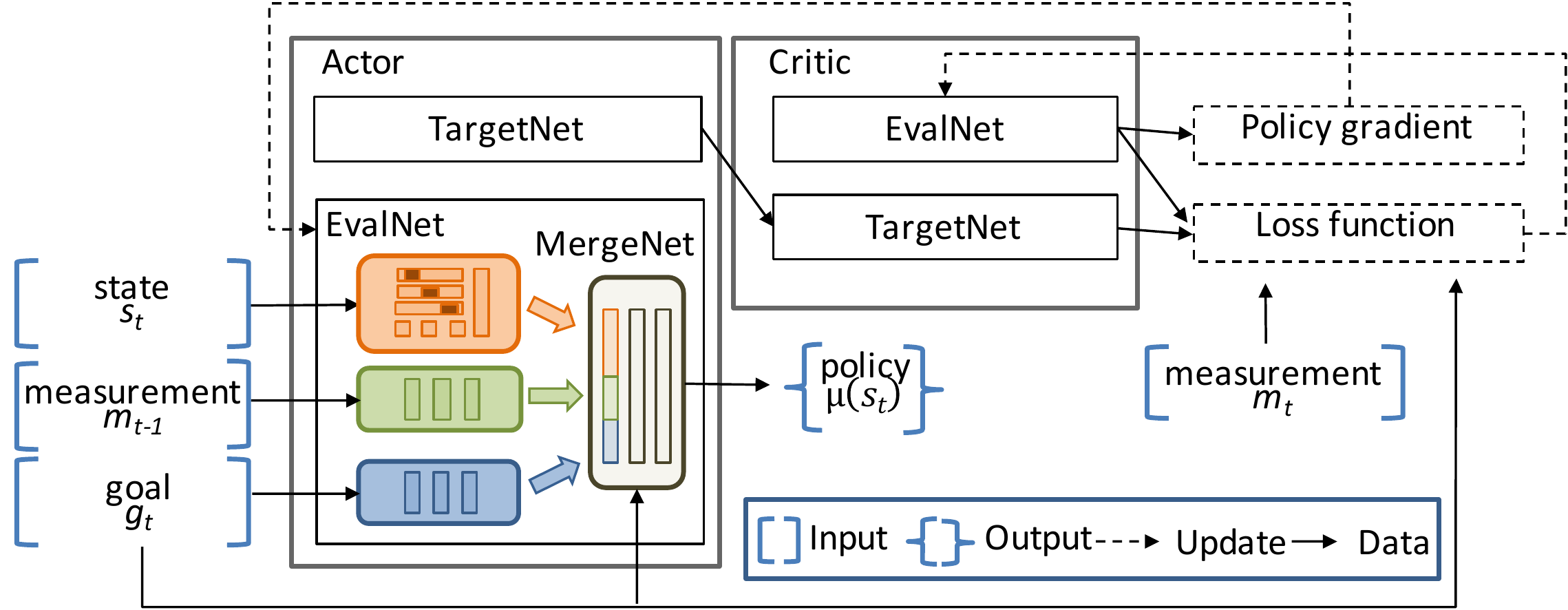}}
			% \vspace{-0.1in}
			\caption{\label{training workflow} Multi-objective DDPG optimization.}
% 			\vspace{-0.15in}
	    \end{figure}

    % \subsubsection{Neural Network Architecture}
	\noindent\textbf{Neural network architecture. }
	% Unlike the tabular learning approach, DeepCC agent takes the high expressiveness of DNNs to handle all input data. 
	We design a multi-objective DDPG algorithm, which involves four deep NNs and exploits the actor-critic algorithm to train the policies on continuous action space. In contrast to basic DDPG architecture, as shown in Fig.\ref{training workflow}, not only the measurement and goal as inputs are added to the agent, but the goal vector is added to each layer of NNs. These explicitly added inputs increase the sensitivity of decision policy to different optimization objectives.

	% In detail,  Actor networks are responsible for choosing the proper action. Critic networks estimate the value of an action and conduct to update the parameters of actor and critic networks.
	% Benefiting from the advances of double deep Q-learning (DQN)~\cite{Hasselt2015DQN}, actor and critic networks are separately presented by two neural networks, as shown in Fig. \ref{training workflow}. The actor networks are divided into an evaluation network (termed \emph{evalNet}) and a target network (termed \emph{targetNet}). 
	% % The targetNet in actor is responsible for updating critic network. 
	% % The evalNet in actor outputs the action from the merge network (termed \emph{mergeNet}), which merges the state,  measurement and goal networks. 
	% The critic networks are also divided into a \emph{targetNet} and an \emph{evalNet}. They have similar outputs with respect to the value of current state, but their input are different. 
	% % The targetNet in critic takes action from the targetNet of actor, while the evalNet in critic takes the action from the evalNet of actor. 
	% In brief, all the neural networks are trained offline, but only the evalNet of actor will be responsible for making decision in the online phase.
% \vspace{-0.06in}    

    % \subsubsection{Training with Multi-Objective Optimization}
    \noindent\textbf{Training with multi-objective optimization. }
	\label{design-training}
		We now describe the policy gradient training algorithm of DeepCC. The algorithm uses two  evaluation networks (termed \emph{evalNet}) to approximate the actor function $\mu(s,m,g|\theta^{\mu})$\footnote{In the following, we use $\mu(s|\theta^{\mu})$ as a shorthand when no ambiguity exists.} and the critic function $Q^{\mu}(s,g,a|\theta^{Q})$, respectively. The two target networks (termed \emph{targetNet}), $\mu(s|\theta^{\mu^{\prime}})$ and $Q^{\mu}(s,g,a|\theta^{Q^{\prime}})$ are the copies of the evalNets accordingly.
		During training, only the two evalNets are trained in each time step, while the parameters of targetNets are updated by slowly tracking the evalNets: $\theta^{\prime} \leftarrow \tau\theta + (1-\tau)\theta^{\prime}$ with $\tau \ll 1$. 
		The critic function can be described in a recursive expression:
% 		\vspace{-0.05in}
		\begin{equation}
			Q^{\mu}(s_{t},g_{t},a_{t}|\theta^{\mu})=\mathbb{E}[r_{t}+\gamma Q^{\mu}(s_{t+1},g_{t+1},\mu(s_{t+1}))] \label{eq6}
% 		\vspace{-0.03in}
		\end{equation}
		where $\gamma$ is the discount factor and $r_{t}$ is the instant reward which is expressed as the combination of $m_{t}$ and $g_{t}$.
		% measurement and goal according to \S\ref{design-multiobj}. 
		Recall from the definition of action space, the actor function $\mu(s|\theta^{\mu})$ specifies the current policy by deterministically mapping inputs to a specific action. It can be updated by applying the chain rule to the expected cumulative reward $J_{mul}$ with a refection to the actor parameters $\theta^{\mu}$:
% 		\vspace{-0.1in}
% 		\begin{small}
			\begin{equation}
			\nabla_{\theta^{\mu}}J_{mul}\approx \frac{1}{N} \sum_{t}\nabla_{a}Q(s_{t},g_{t},a|\theta^{Q})|_{a=\mu(s_{t})}\nabla_{\theta^{\mu}}\mu(s_{t},g_{t}|\theta^{\mu}) \label{eq7}
		  %  \vspace{-0.03in}
			\end{equation}
% 		\end{small}
		where $N$ is the training batch size. 

		The objective of the training is to maximize $J_{mul}$ and minimize the loss $L$ of critic network, which is defined as:
			\begin{equation}
			L=\frac{1}{N}\sum_{t}(y_{t}-Q(s_{t},g_{t},a_{t}|\theta^{Q}))^{2} \label{eq9}
% 			\vspace{-0.03in}
			\end{equation}
		where $y_{t}$ is estimated by:
			\begin{equation}
			y_{t}=r_{t}+\gamma Q^{\prime}(s_{t+1},g_{t+1},\mu^{\prime}(s_{t+1}|\theta^{\mu^{\prime}})|\theta^{Q^{\prime}}) \label{eq10}
% 			\vspace{-0.03in}
			\end{equation}
		To achieve the multi-objective training, we randomly set the goal vector $g_{t}$ for each training episode. Therefore, the DeepCC could successfully generalize across a wide range of trade-offs between different performance metrics.
        When $L$ tends to zero and the average $J_{mul}$ of a batch size data no longer increases, the agent is considered to converge.
        			
		As a summary, the actor and critic function are trained by the policy gradient and value-based method respectively.
		After training, the actor function parameterized by $\theta^{\mu}$ will converge to different near-optimal policies corresponding to different objectives with the help of the critic function. Once being trained, the agent learns near Pareto front with different goals under different network conditions (See  \S\ref{offline-evaluation}). 
	\subsection{Online DeepCC tuning}
	\label{design-online}
    % \begin{figure}[t] 
 	% 		\centerline{\includegraphics[width=0.7\linewidth]{figures/online-architecture.pdf}}
	% 		\vspace{-0.15in}
    %         \caption{\label{online-architecture} Online learning.}
	% 		\vspace{-0.25in}
	% \end{figure}
	
	% 	To reach the global optimum of online tuning algorithm,
	
	How to adaptively choose the proper policy form learned Pareto optimal policies remains a challenge. To address this issue, we design a policy controller, i.e., an online tuning module to achieve the requirement about throughput, delay, and loss rate, i.e., the \emph{target}. Our key insight is that we can regard the action regulation as a black box where the input includes  state, measurement, and goal driven by the ACK signal. Then the \emph{goal} performs as a knob that controls the agent to pursue different targets.
	Specifically, DeepCC continuously detects the changes in the difference between current measured performance and application requirements, and automatically tunes the \emph{goal} with multi-dimensional gradient descent.
% 	optimize for the different objectives.
    
	% As shown in Fig.~\ref{online-architecture}, 
	During a TCP session, we adopt the multi-dimensional gradient descent algorithm to tune the agent online. According to the changes in network conditions and the prior measurements, the multi-dimensional gradients about the distance between measurement and target value are computed and applied to generate a new \emph{goal} for the agent.  
	Thereby the current measurement can be fed to enable a negative-feedback loop to influence the action choice until the performance converges to the application requirements under unseen network conditions. 

	\noindent\textbf{Multi-dimensional gradient descent algorithm. }
% 	The  agent is generated by offline learning and its objective function (controlled by the \emph{goal}) can be tuned in online stage according to the \emph{target} of applications. 
% 	We design an online tuning algorithm based on multi-dimensional gradient descent. Our online tuning algorithm aims to reduce the distance between the current measurements and the target values.   
    The online tuning algorithm is based on multi-dimensional gradient descent.
	Specifically, the objective of the online tuning algorithm is to minimize the loss function $J=Euclid(m_{t},T)$ with respect to the goal $g_{t}$. Here, (1) $m_{t}$ represents the performance under the corresponding goal $g_{t}$ at time step $t$. In this way, we suppose that the measurement can be regarded as a function of $g_{t}$, i.e., $m_{t} := f(g_{t})$; (2) the $n$-dimensional target value $T=(T^{(1)}, T^{(2)}, \cdots ,T^{(n)})$ is specified by the application;(3) $Euclid(m_{t},T)$ represents the euclidean distance between the measurements and the target value. 

	To reduce the variance of the estimated value $m_{t}$, we set a k-step average of $J$ as the loss function in practice. Taking the relationship between $m_{t}$ and $g_{t}$ into account, the actual loss function is described as:
	\begin{equation}
		J(g_{t}) = \frac{1}{2k}\sum_{i=t-k}^{t}(Euclid(f(g_{t}),T))
% 	\vspace{-0.03in}
	\end{equation}
	
	In this paper, the target value $T=(T^{(1)},T^{(2)},T^{(3)})$ is defined as the desired max-min bound for each dimensional performance metric.
	The detailed descriptions of $T^{(i)}$ are shown in Table \ref{Notation}.
% 	i.e., $M^{(1)}$ is the minimum throughput, $M^{(2)}$ is the maximum delay and $M^{(3)}$ is the highest packets loss rate. 
	In practice, if $m^{(i)}$ is satisfied the bound of $T^{(i)}$, we set $f(g_{t}^{(i)})=T^{(i)}$. 
	Namely, we only update gradients when the target requirements are not satisfied. 
	If the application can not provide the concrete performance target but have a performance preference, e.g., the high throughput or low delay, the target $T$ could be presented as an ($n$-1)-dimensional vector with the default value. In this case, the target $T$ is not related to the uninterested metric. Under the low-delay mode, the target $T$ is set as the minimum RTT and zero packet loss. While in the high-throughput mode, the target $T$ can be defined as the observed maximum bandwidth and zero packet loss. 

	We use a gradient descent method to update the $g_{t}$. 
	The gradient of $J$ with respect to $g_{t}$ can be derived as:
	\begin{equation}
		\bigtriangledown_{g_{t}}J(g_{t}) = \bigtriangledown_{f(g_{t})}J(f(g_{t}))\bigtriangledown_{g_{t}}f(g_{t}) 
% 	\vspace{-0.03in}
	\end{equation}
	
	Since it is non-trivial to model the relationship between the goal $g_{t}$ and the measurement $m_{t}$, we can not easily obtain the analytic expression of the function $f$. Here we use a numerical method to approximate $\bigtriangledown_{g_{t}}f(g_{t})$:
    \begin{equation}
    		\bigtriangledown_{g_{t}}f(g_{t}) \approx \frac{1}{k}\sum_{i=t-k}^{t-1}\frac{\Delta m_{i} }{\Delta g_{i}} =\frac{1}{k}\sum_{i=t-k}^{t-1} \frac{ m_{i+1}-m_{i} }{g_{i+1}-g_{i}}
    % \vspace{-0.03in}
    \end{equation}
    
   Finally, the goal can be updated by the following formula.
	\begin{equation}
		g_{t+1} \gets g_{t} - \alpha*\bigtriangledown_{g_{t}}J(g_{t})
		\label{update_gradient_equation}
% 	\vspace{-0.03in}
	\end{equation}
	where $\alpha$ is the learning rate.

	Since the specific function form of $f(g_{t})$ is unknown and the function is not necessarily a convex function, the traditional gradient descent method may fall into local optimum, resulting in performance degradation. To solve this problem, we design a simple rule-based approach to help ``jump out" of the local optimum. In practical terms, if the distance between the $i$th-dimensional measurement and target is larger than a threshold, the $i$th-dimensional goal is increased by a pre-setting value.

	Moreover, non-congestion packet loss is a common phenomenon in the Internet~\cite{Dong2018Vivace}. The goal of loss rate (i.e., $g^{(3)}$) should not change due to random packet loss.
	For DeepCC, we set the constraint for the gradient of the loss rate. 
	That is to say, if the target is not satisfied with the measurement of throughput or delay, the gradient of loss rate will be set to equal that of delay.
    The above method ensures that the gradient of loss is not affected by non-congestion packet loss since it always follows the gradient of delay.

	Since the sum of each dimension of goal vector is 1 in the training process, each dimension of updated goal $g^{(i)}$ will be normalized before fed into the agent as follow:
	\begin{equation}  
		\label{normalize g}
		g^{(i)} = \frac{g^{(i)}}{\sum_{i=1}^{n}g^{(i)}}
	\end{equation}
    
	In fact, it could cause a large variance of goal and huge fluctuations in the convergence process when updating $g_{t}$ directly following e.q. \eqref{update_gradient_equation}. Additionally, it is difficult to choose the proper learning rate $\alpha$. An adaptive learning rate algorithm Adam~\cite{Kingma2014Adam} is a first-order optimization algorithm that is less sensitive to the choice of learning rate than the basic gradient descent algorithm. It also has advantages in non-convex optimization problems. Therefore, we combine Adam algorithm to update the gradient $\bigtriangledown_{g_{t}}f(g_{t})$ in practice.

\section{Implementation and Training}
    \label{sec:impl}
    % To implement DeepCC, we build a user-space implementation and the DRL architecture of DeepCC is implemented with TensorFlow~\cite{TensorFlow}.
    % We have implemented DeepCC in user-space as the congestion control module that the endpoints can communicate with over UDP. 
    In this section, we describe the DeepCC implementation, training and the interface.

    \noindent\textbf{DeepCC implementation.}
        \label{impl:DeepCC implementation}
	We implement the sender and receiver in the user space by adopting the UDP-based transmission skeleton, like Indigo\cite{Francis2018Pantheon}. The sending and receiving events are implemented by the message-triggered mechanism. DeepCC replaces the sender-side congestion control with the offline-learned agent and the online tuning module. The sender program receives the targets from the applications, executes the action, and controls the packets sending behavior. Unlike the sender, DeepCC remains the receiver unchanged. 
    % DeepCC uses the user-space implementation based on UDP. 

    For practical implementation, the sender firstly loads the multi-objective agent and sets the default \emph{goal} according to the target. 
    Every time step the sender received an ACK message, it updates the estimated measurement and state. Then it infers the next-step sending rate via the agent with the \emph{goal}, measurement, and state. Once getting the sending rate, the sender can calculate the new \emph{cwnd} size and pace these packets in an ack-clock.  
    In the process of sending, if the current throughput and delay do not reach the target, the sender tunes the \emph{goal} according to the online tuning algorithm. 
    % presented in \S\ref{design-online}.

% \subsection{Decision interval }
%     \label{impl:decision-interval}
% The overhead of DeepCC mainly lies in obtaining the action and updating the \emph{goal}, since the neural networks and the online tuning algorithm introduce extra complexity at the endpoint. 
    The overhead of DeepCC mainly lies in obtaining the action and updating the goal, since the neural networks and the online tuning algorithm introduce extra complexity at the endpoint. To balance the overhead and efficiency, DeepCC triggers the model inference and goal tuning every interval instead of at the packet level. Specifically, the model inference and goal tuning are triggered when an ACK is received and the time since the last decision exceeds the decision interval, e.g., 1/2 RTT and 4 RTT respectively. 
    % As a naive approach, the sender makes decision-making when receiving an ACK. However, the model inference and goal tuning could block the sender and impact the performance in the real environments. To address the overhead of model inference, DeepCC balances the overhead and the efficiency by following intervals about action inference and goal tuning. 
    % The frequency of model inference and goal tuning impacts the overhead. To balance the overhead and efficiency, 
    % DeepCC triggers the model inference and goal tuning every interval instead of at the packet level. Specifically, the model inference and goal tuning are triggered when an ACK is received and the time since the last decision exceeds a (different) decision interval. Typically, we set the model inference and goal tuning interval as 1/2 RTT and 4 RTT respectively(\S~\ref{eval:overhead}).
    Furthermore, DeepCC takes an asynchronous interaction between the model inference, goal tuning, and packet sending so that they do not need to wait for other executions.
    % \zl{However, how to properly set the decision interval becomes a critical problem.  On one hand, if the decision interval is too large, it can not track the changes in network conditions in time, thereby fail to react to the changes in network states. On the other hand, if the decision interval is too small, the number of decision-making will increase resulting in high the inference frequency. Therefore, }
    
 {\noindent\textbf{Training. }}
    We implement and train the multi-objective agent in Python with Tensorflow\cite{TensorFlow} for ease of development.
    % \noindent\textbf{Neural network architecture.} 
	To learn the control policies in the actor network, DeepCC first separately extracts the feature from the three inputs, i.e., state, measurement, and goal. The three inputs employ different neural network structures. States with 16 past values are passed to a 1D convolution neural network with 32 filters, each of size 3 with stride 1. The measurement and goal networks use three fully connected layers. 
	Then mergeNet takes all the processed features from the above three networks and outputs the sending rate with the ``tanh" activation function. The critic network takes the state, goal, and action as inputs and uses the same architecture as the actor network to conduct feature extraction. The extracted features of action are also processed by three fully connected layers, which is then fed to the last layer of the critic network. The difference between the actor and critic networks is that the final output of critic network is a linear neuron without activation function. 
	
	The agent is trained under the network environment where the fluctuation of bandwidth follows Poisson distribution, like the Wi-Fi link (See \S\ref{offline-evaluation}).  Once the multi-objective agent is well trained offline, this agent will not be retrained in practical deployment.
    % 	Note that this offline training time is measured only with the CPU enabled and it could be further sped up using GPU.  
    % 0.5$\sim$1.3
	
	% During the online stage, the inference time of multi-objective agent takes about average 0.9 ms to get the {\em{action}} or update the {\emph{goal}}. 	
	% \zl{Figure. \ref{fig:online-decision} shows the throughput of DeepCC under different decision makings.}
	
    % \zl{To reduce the overhead of these operations, we set the decision interval to obtain the action from the agent.  }
	% and the goal-tuning interval as 4 minimum RTT.	
    \noindent\textbf{Interface.}
    \label{impl:API}
    In order to satisfy diverse requirements, DeepCC provides the interface for applications to set their specified performance targets or preferences. This allows CC to directly perceive application requirements. The interface takes two forms. On one hand, applications can specify the target value of their required throughput, delay, and packet loss rate. The performance requirements provide optimization objectives for the online tuning algorithm. On the other hand, if the applications cannot provide the value of requirements, they can choose their performance preferences through the interface with the second form. In this case, the application can choose the high-throughput mode or low-latency mode (\S\ref{design-online}).

\begin{figure*}[t]
        \centering
        % \vspace{-0.15in}
        \subfigure[Offline performance over Wi-Fi link]{
            \label{Dynamic-offline} %% label for first subfigure
            \includegraphics[width=0.32\linewidth]{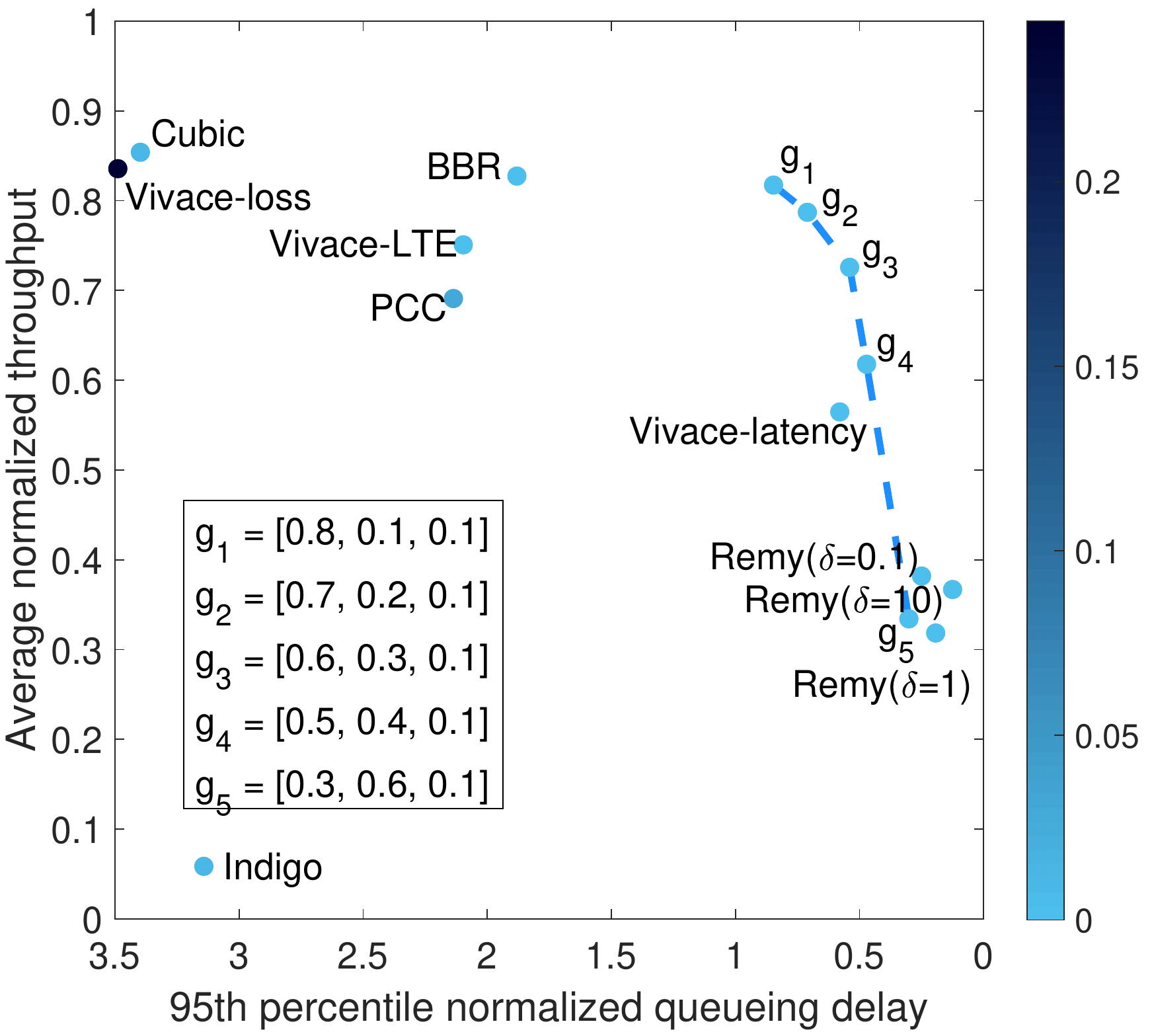}}
        % \hspace{0.8in}
        % \hspace{0.01in}
        % \vspace{-0.1in}
        \subfigure[Offline performance over cellular link]{
            \label{LTE-offline} %% label for second subfigure
            \includegraphics[width=0.32\linewidth]{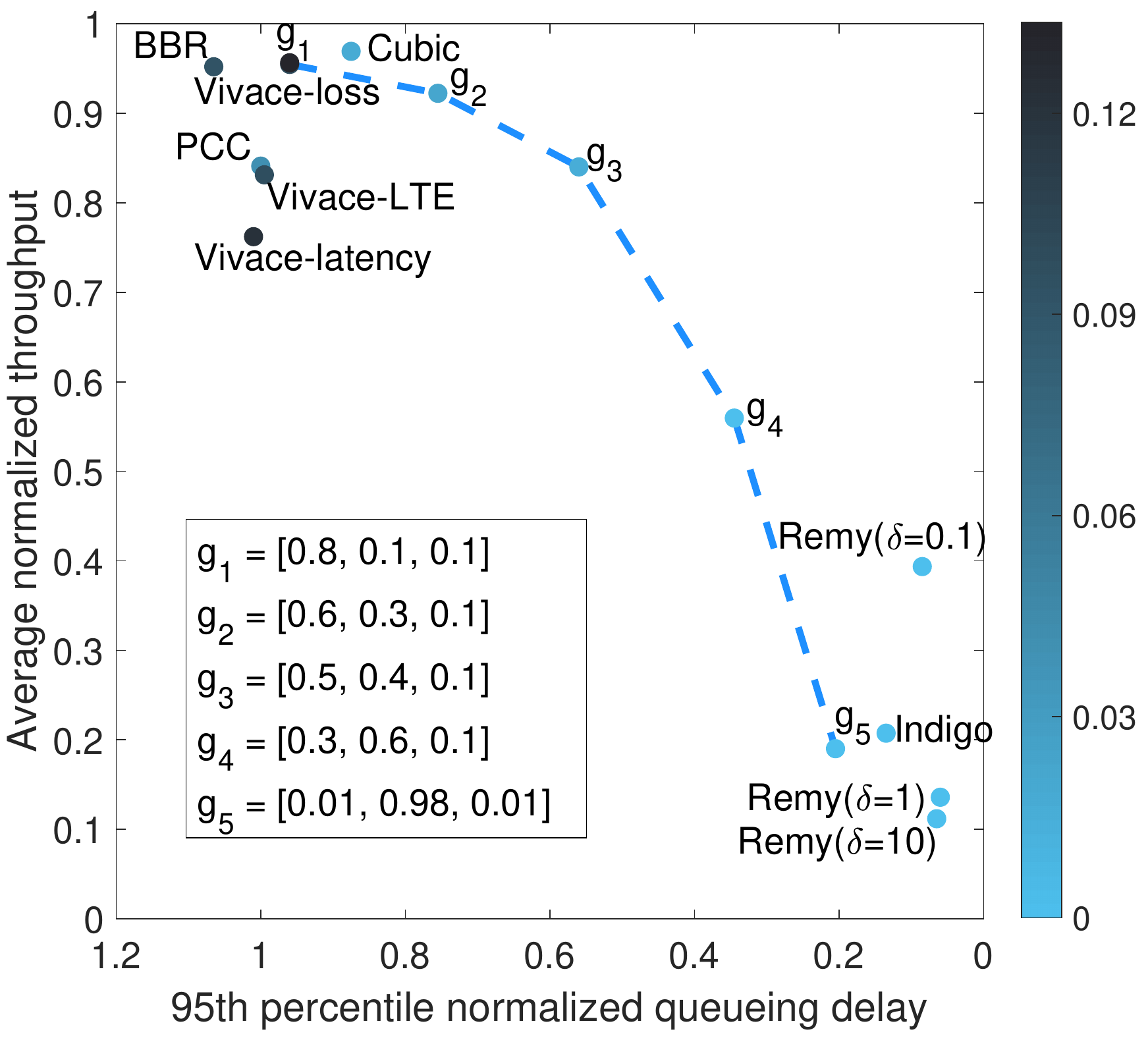}}
        % \hspace{0.01in}
        \subfigure[Offline performance over satellite link]{
            \label{fig:offline-satellite} %% label for second subfigure
            \includegraphics[width=0.32\linewidth]{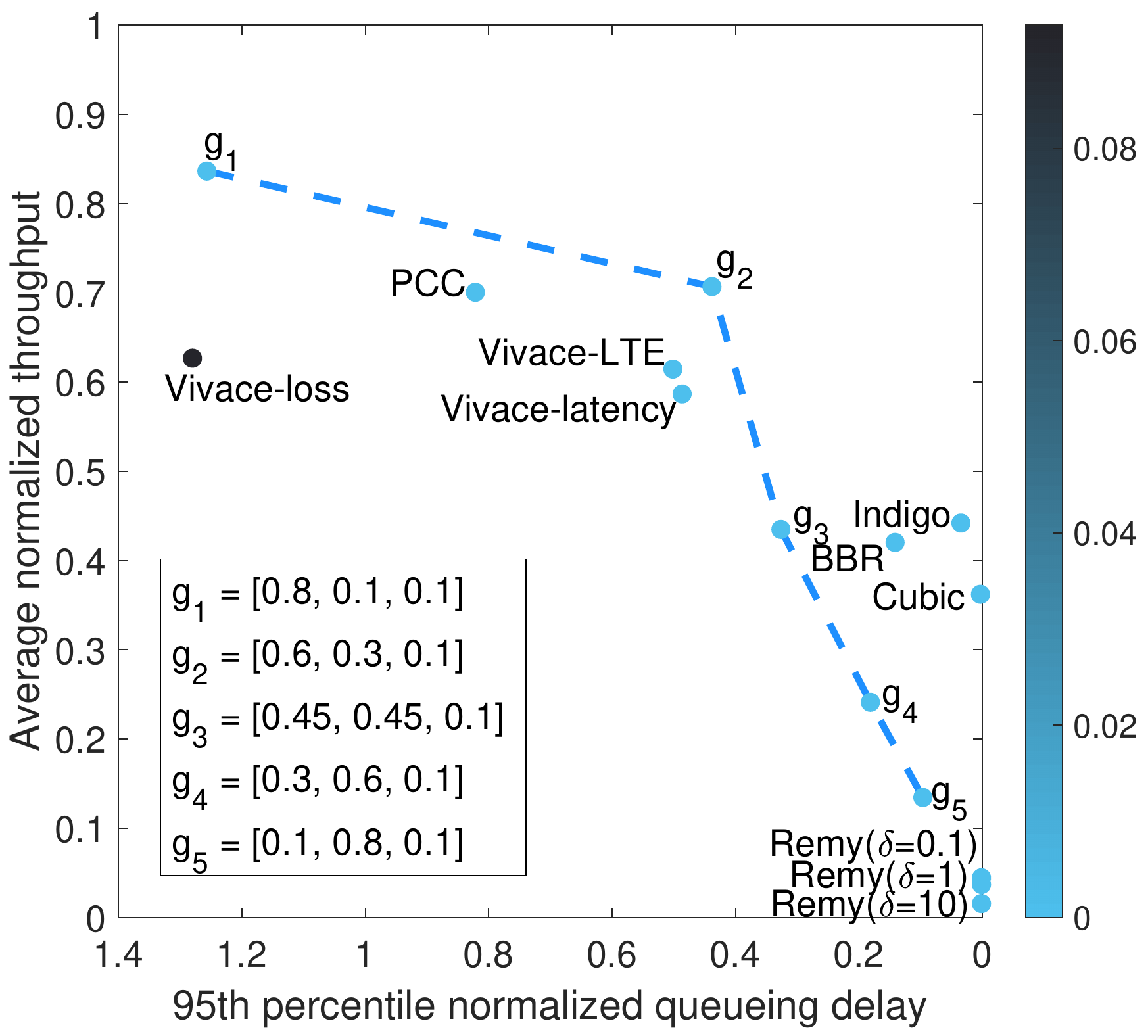}}
    % 		\hspace{-0.1in}
        % \vspace{-0.1in}
        \caption{Performance frontier achieved by {offline learned DeepCC agent}. The color bar represents the packet loss rate where darker is worse. }
        \label{offline-fig} % % label for entire figure
        % \vspace{-0.15in}
\end{figure*}
\begin{figure*}[t]
        \centering
        \subfigure[Online performance over cellular link]{
            \label{ATT-online}
			\includegraphics[width=0.32\linewidth]{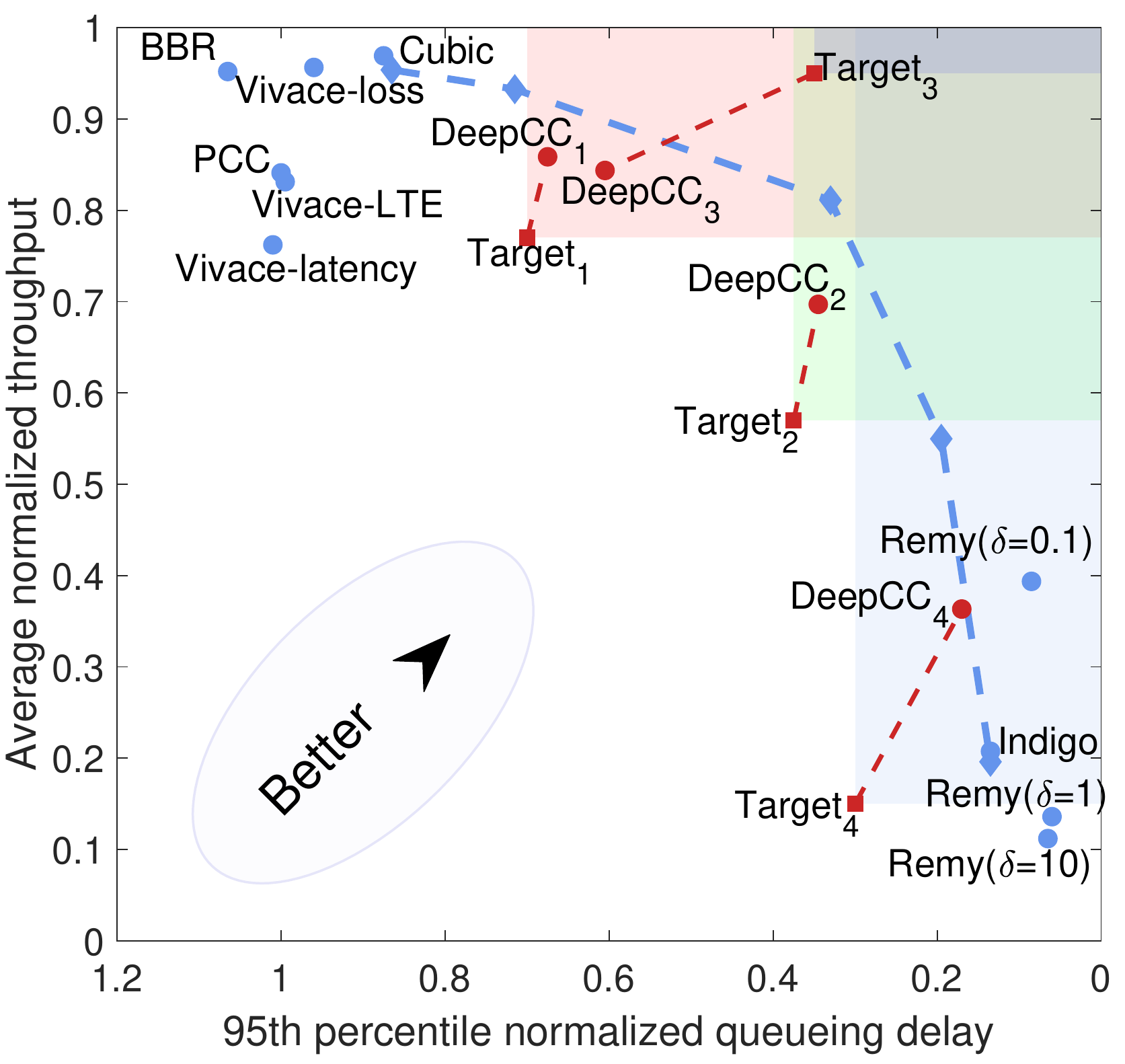}}
        \subfigure[Inter-Continental link (CN-USA)]{
            \label{fig:realworld}
            \includegraphics[width=0.32\linewidth]{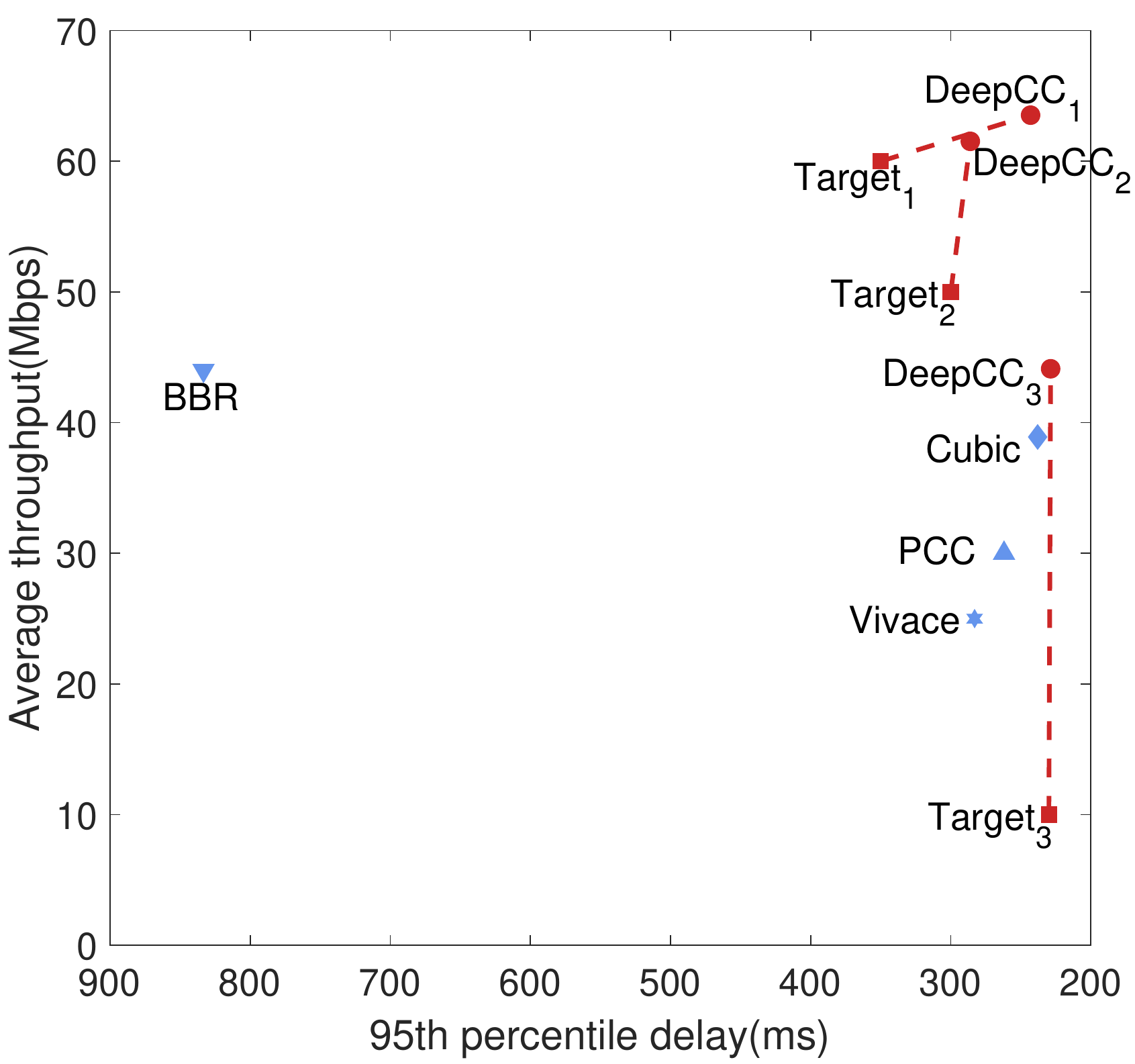}}
        % \hspace{0.01in}
        \subfigure[Intra-Continental link (CN-JPN)]{
            \label{fig:Aliyun}
            \includegraphics[width=0.32\linewidth]{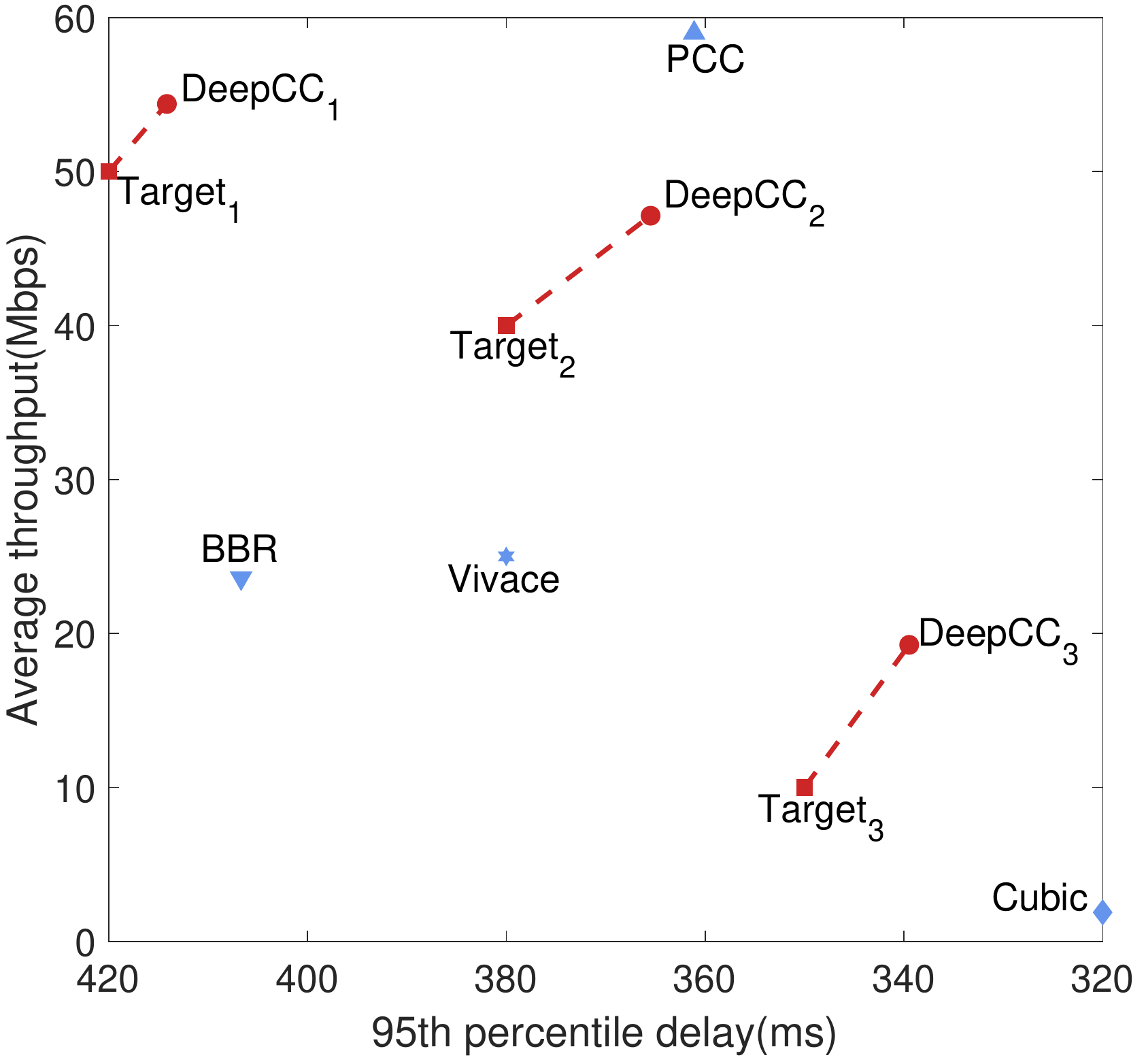}}
		\caption{Online performance frontier achieved by DeepCC. {The latter two experiments are conducted over real-world network links.}}
% 		\vspace{-0.15in}
\end{figure*}

\section{Evaluation}
	\label{sec:evaluation}
	{
	In this section, we demonstrate the advantages of DeepCC's multi-objective design in both emulator and real-world scenarios over various state-of-the-art schemes. In particular, we illustrate DeepCC's multi-objective high performance in different networks (\S\ref{offline-evaluation}), high target completion ratio (\S\ref{online-evaluation}), decision interval and overhead (\S\ref{eval:overhead}), friendliness and fairness (\S\ref{eval:fairness}), and robustness (\S\ref{common properties}).
	}

\subsection{Offline behavior over multi-objective optimization}
	\label{offline-evaluation}
		To validate the effectiveness of multi-objective optimization, we repeatedly run DeepCC's agent under three network scenarios with different {\em{goals}} in a broad range using Mahimahi\cite{Netravali2015Mahimahi}. Among them, we select 5 representative results, which are in the mode of high throughput, low delay, or the trade-offs.
% 		We also compare DeepCC with not only PCC, Vivace, Remy and Indigo with their corresponding objective functions, but Linux kernel implementation schemes i.e., Cubic and BBR.
		We also compared DeepCC with the off-the-shelf schemes. As shown in Fig.~\ref{Dynamic-offline} and (b), we intuitively visualize the performance of different schemes in a 2D throughput-delay space. The corresponding loss rate is shown via the color filled.
		% The detailed observations are analyzed as follows.

		\noindent \textbf{Emulated {Wi-Fi} link. }
		First, we test DeepCC over an emulated {Wi-Fi} link (from Pantheon~\cite{Francis2018Pantheon}) with high bandwidth variations, which DeepCC has been trained on. The link has an average 2.64 Mbps bandwidth following Poisson distribution, 176 ms minimum RTT and a drop-tail queue with 130 packets of buffer. As shown in Fig.~\ref{Dynamic-offline}, DeepCC with 5 representative goals achieves an efficient frontier that covers a wide range of trade-offs between throughput, queueing delay, and loss rate.

		Specifically, DeepCC with $g_{1}$ in high-throughput mode achieves comparable throughput with the state-of-the-art CCs, while reduces the queueing delay by up to 75\% with almost zero packet {loss}. When running in the low-latency mode, DeepCC with $g_{5}$ performs almost the same as Remy. Other trade-off points stand in the middle and attain different trade-offs. These results show the efficiency of our multi-objective optimization.
		So DeepCC can achieve a large range of accessible trade-offs only by tuning the goal without any retraining or redesigning efforts and it achieves almost zero packet loss without sacrificing other performance metrics due to the high penalty of loss rate in the reward function.

		\noindent \textbf{Emulated cellular link. }
		To evaluate the performance of DeepCC in an unseen network link, we replay the AT\&T driving trace provided by~\cite{Winstein2013Stochastic}. The cellular link has 200 ms minimum RTT and 140 packets of buffer. As shown in Fig.~\ref{LTE-offline}, the performance of DeepCC with different goals also achieves an efficient frontier.
		This results indicates that through training DeepCC has learned an adaptive policy supporting multi-objective optimization {even under an unseen scenario}. However, to achieve a similar relative position of Fig.~\ref{Dynamic-offline}, DeepCC may require different goals. This phenomenon exposes the challenge on how to set the proper goal according to the network condition, which will be described in the following with our online tuning algorithm.

		\noindent \textbf{Emulated satellite link. }We also evaluate DeepCC'agent on an emulated satellite link with the same setup in PCC\cite{Dong2015PCC} and Copa\cite{Venkat2018Copa} papers. The satellite link has 42 Mbps capacity, 800 ms RTT, 1 BDP (bandwidth-delay product) buffer and 0.74\% stochastic loss rate. Fig. ~\ref{fig:offline-satellite} shows that DeepCC with $g_{1}$ in max-throughput mode obtains higher
		throughput than {that of} other schemes. DeepCC with $g_{5}$ in delay-sensitive mode achieves lowest queueing delay. BBR which performs well in the above two scenarios, can not achieve persistent performance in this scenario. Furthermore, DeepCC with $g_{2}$ outperforms PCC and Vivace with much higher throughput and lower queueing delay.
		
		{{As shown in} the above experiments, DeepCC can {reach a wide range of performance by flexibly adjusting the goal}. DeepCC can improve the throughput from a low-throughput point to a high-throughput {one} up to 9X. Likewise, it can reduce the queueing delay from a high-latency point to a low-latency {one} {up} to 10X.}
		
% 		But the generalization ability to adapt to different network conditions cannot be fully guaranteed when DeepCC only involves offline training.

\subsection{Online performance over specified requirements}
	\label{online-evaluation}
	\noindent \textbf{Emulated cellular link. }
	As shown in Fig.~\ref{ATT-online}, {DeepCC gains an efficient performance frontier} again in the online stage. Compared with the frontier achieved by offline learning shown in Fig.~\ref{LTE-offline}, DeepCC with online tuning achieves a wider frontier than purely through offline learning. That is because DeepCC takes advantage of both offline and online learning.
    When running online, it can timely regulate the near-optimal policy based on the offline trained agent according to the real-time network conditions.
	To provide a zoom-in view of online tuning, {we carefully select four targets, among whom three are achievable (i.e., inside the frontier) and one is unachievable (i.e., outside the frontier). The shadow area of each achievable target represent the satisfactory region of the corresponding application. Note that the region under the frontier of three achievable targets are non-overlap so that they can not be satisfied with DeepCC with a single goal. DeepCC achieves them with proper goal-tuning through its online learning algorithm. For the unreachable target,} DeepCC tries to achieve the performance close to the frontier.

% 	we select four targets and the corresponding achieved performance.
	% Some of the targets are achievable while some are not.
    % As shown, DeepCC tries to achieve the performance close to the frontier even if the target is unreachable.

	\noindent \textbf{Evaluation in the wild. }
		\label{eval:wild}
		% We evaluate DeepCC in a realistic environment. The sender is deployed in Aliyun\cite{aliyun}. The receiver is located inside our campus and connected to the sender through the wide-area network. To validate the efficiency of DeepCC, we use tc\cite{TC} to add extra 62 KB buffer on this path (See \S\ref{discussion}). The resulting average bandwidth and minimum RTT of this path are 0.5 Mbps and 107 ms respectively.
		We evaluate DeepCC over wide-area Internet paths spanning two continents.
		{The senders are located in an Amazon\cite{amazon} cloud server in USA and a server located in Japan.} The receiver is located inside our campus and connected to the senders through the wide-area network.
		% We evaluate DeepCC under three \emph{goals} by only using the offline-learned agent. Further,
		We evaluate DeepCC with three \emph{targets} and compare them against PCC, Vivace, Cubic and BBR~\footnote{Cubic and BBR are implemented in Linux kernel 4.15.0.}.
		The results are shown in Fig. \ref{fig:realworld} and Fig. \ref{fig:Aliyun}. The average throughput and 95th percentile delay achieved by DeepCC are significantly different with three requirements.  Guided by the \emph{target} value, DeepCC can not only satisfy the corresponding requirement but achieve better performance in both throughput and delay. The {delay of DeepCC with low-latency targets achieve} significantly lower than that of BBR, PCC and Vivace. Notice that BBR, PCC and high-throughput DeepCC are not sensitive to packets loss, {so that they both} do well on throughput but high-throughput DeepCC achieves lower delay than that of BBR. 
% 		Cubic is sensitive to packet loss and hence loses throughput in intra-continental link.
\begin{figure*}[t]
	    \centering
	    \subfigure[The target completion ratio over {trained} network link.]{
            \label{fig:online-Wifi-network} %% label for first subfigure
			\includegraphics[width=0.35\textheight]{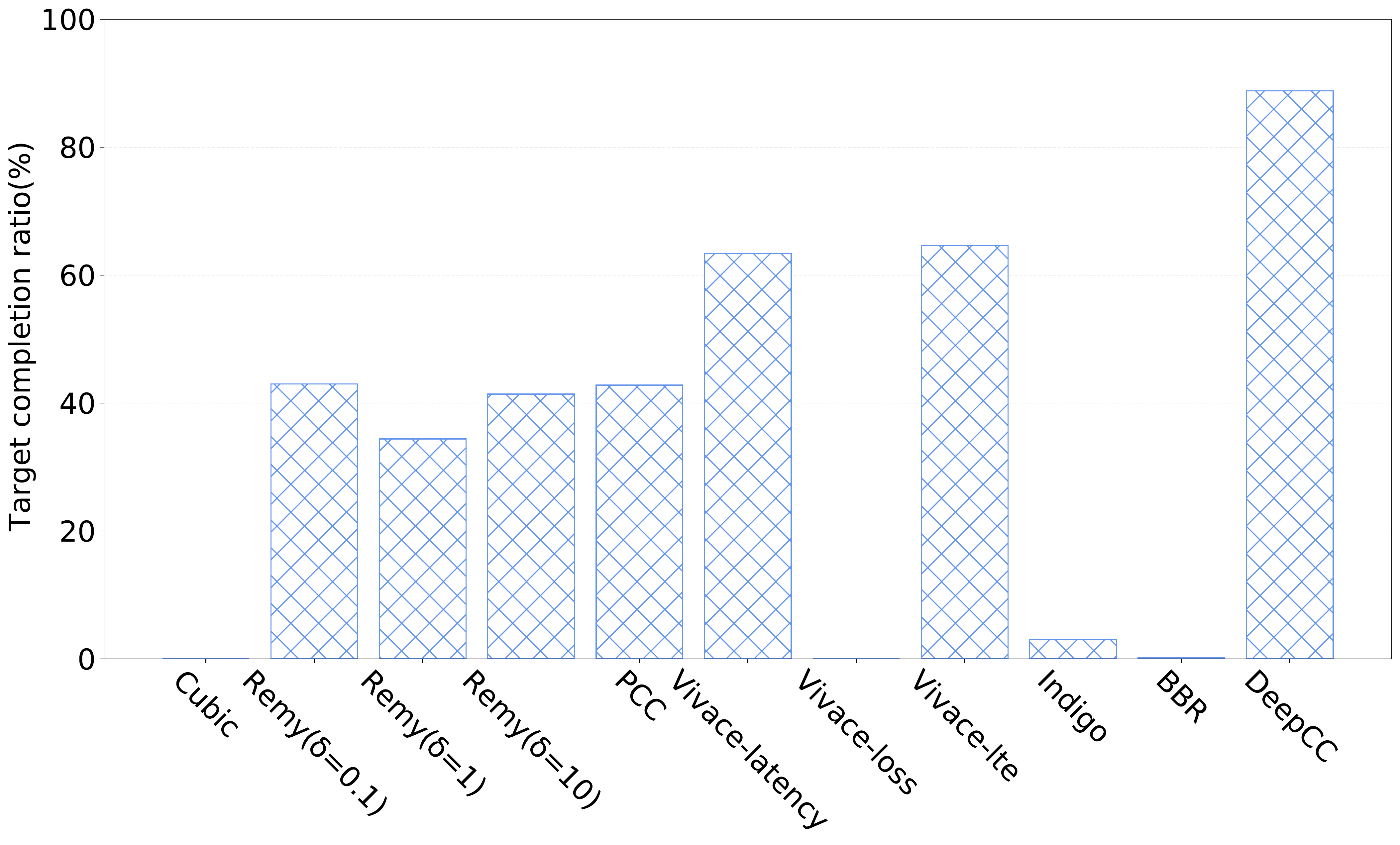}}
		\subfigure[The target completion ratio over {untrained} network link.]{
            \label{fig:online-Cellular-network} %% label for first subfigure
			\includegraphics[width=0.35\textheight]{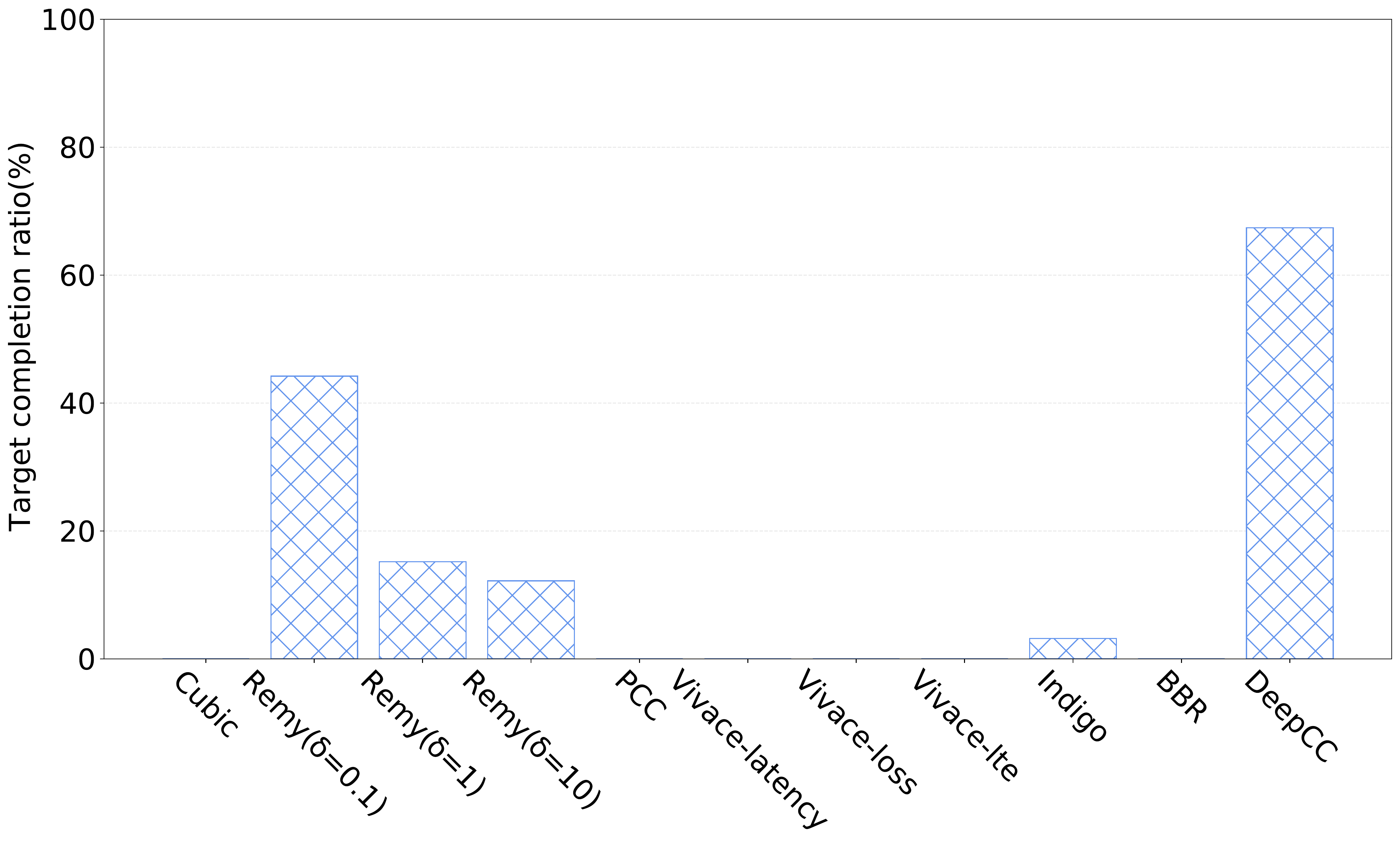}}
		\caption{The target completion rate under 500 random targets of different congestion controls.}
% 		\vspace{-0.15in}
\end{figure*}
	\noindent \textbf{Target completion ratio. }
	To investigate the generalization ability of DeepCC in online stage, we randomly sample 500 target values $T$ of application requirements over above-mentioned {Wi-Fi} link and cellular link. The range of these target values $T$ is selected as follows: $T^{(1)}$ ranges from 0 to the achievable maximum throughput of all {learning}-based schemes on the corresponding link; $T^{(2)}$ varies from 0 to the achieved maximum queueing delay of all schemes on the corresponding link; and $T^{(3)}$, representing the loss rate, ranges from 0 to 10\%. Obviously, {some of them can not be achieved by any one of the existing schemes.}

	Here, we denote the \emph{target completion ratio} (TCR) as the ratio of target achieved to all targets as our quantitative evaluation metric. Intuitively, for a given congestion control scheme, the higher completion ratio is obtained, the more application requirements can be achieved. Since all the existing learning-based schemes lack tunable objectives, we test each scheme with its default objective under the given network link and calculate its completion ratio. For DeepCC, we test it with different targets and leave the agent unchanged.
	The results are shown in Fig. \ref{fig:online-Wifi-network} and Fig. \ref{fig:online-Cellular-network}.
	As expected, DeepCC achieves the highest target completion ratio {among all the schemes}. Even in an untrained network scenario, DeepCC still achieves 67.4\% completion ratio. The results indicate that DeepCC can tune the objective function to adapt to both network conditions and application requirements through adjusting the \emph{goal} .

\begin{figure*}[t]
        \begin{minipage}[t]{0.48\linewidth}
		\centering
		\includegraphics[width=0.99\linewidth]{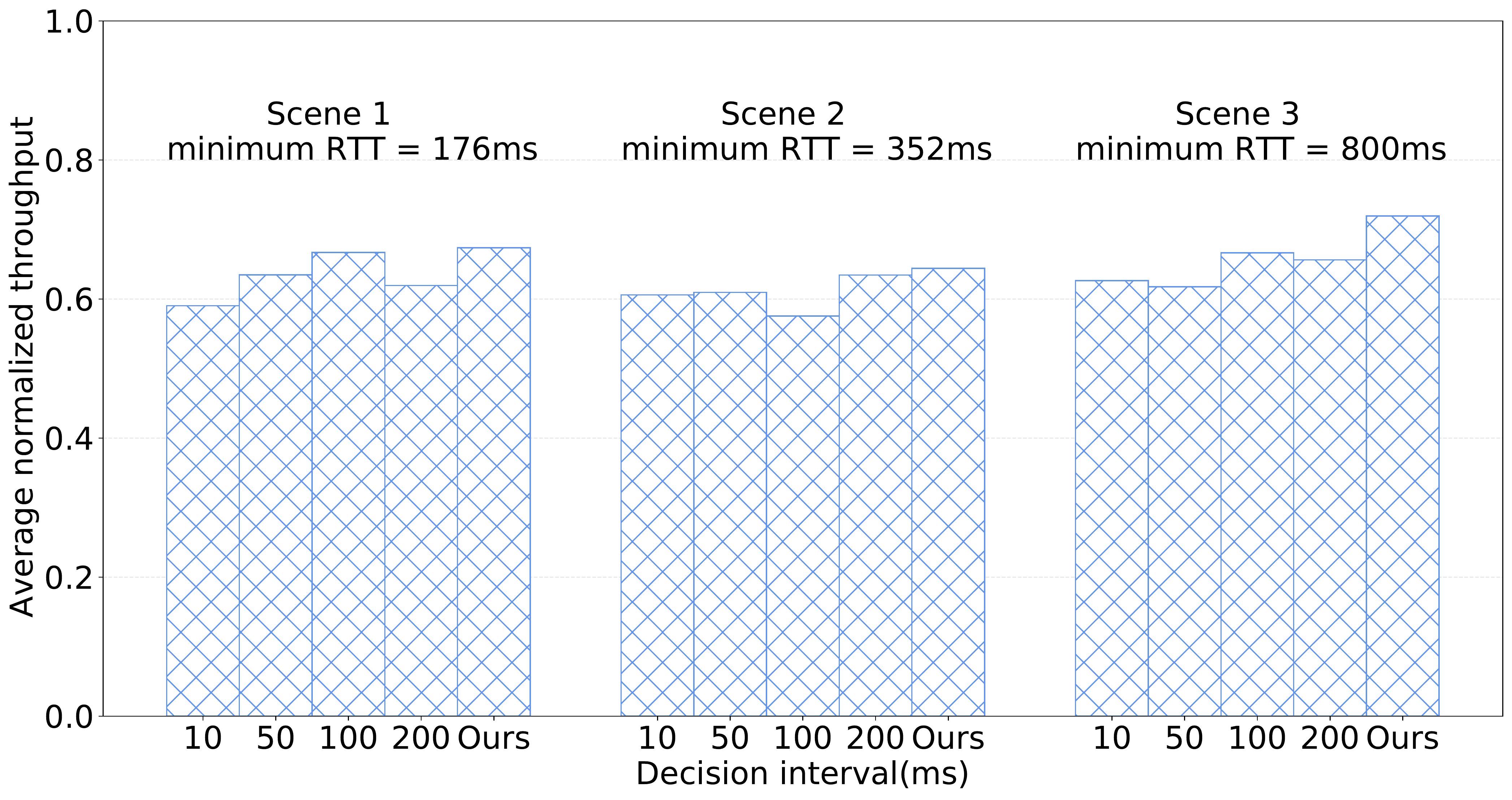}
		\caption{\label{fig:offline-decision}The average normalized throughput across different decision intervals.}
	    \end{minipage}
	    \vspace{-0.1in}
		\begin{minipage}[t]{0.48\linewidth}
		\centering
		\includegraphics[width=0.99\linewidth]{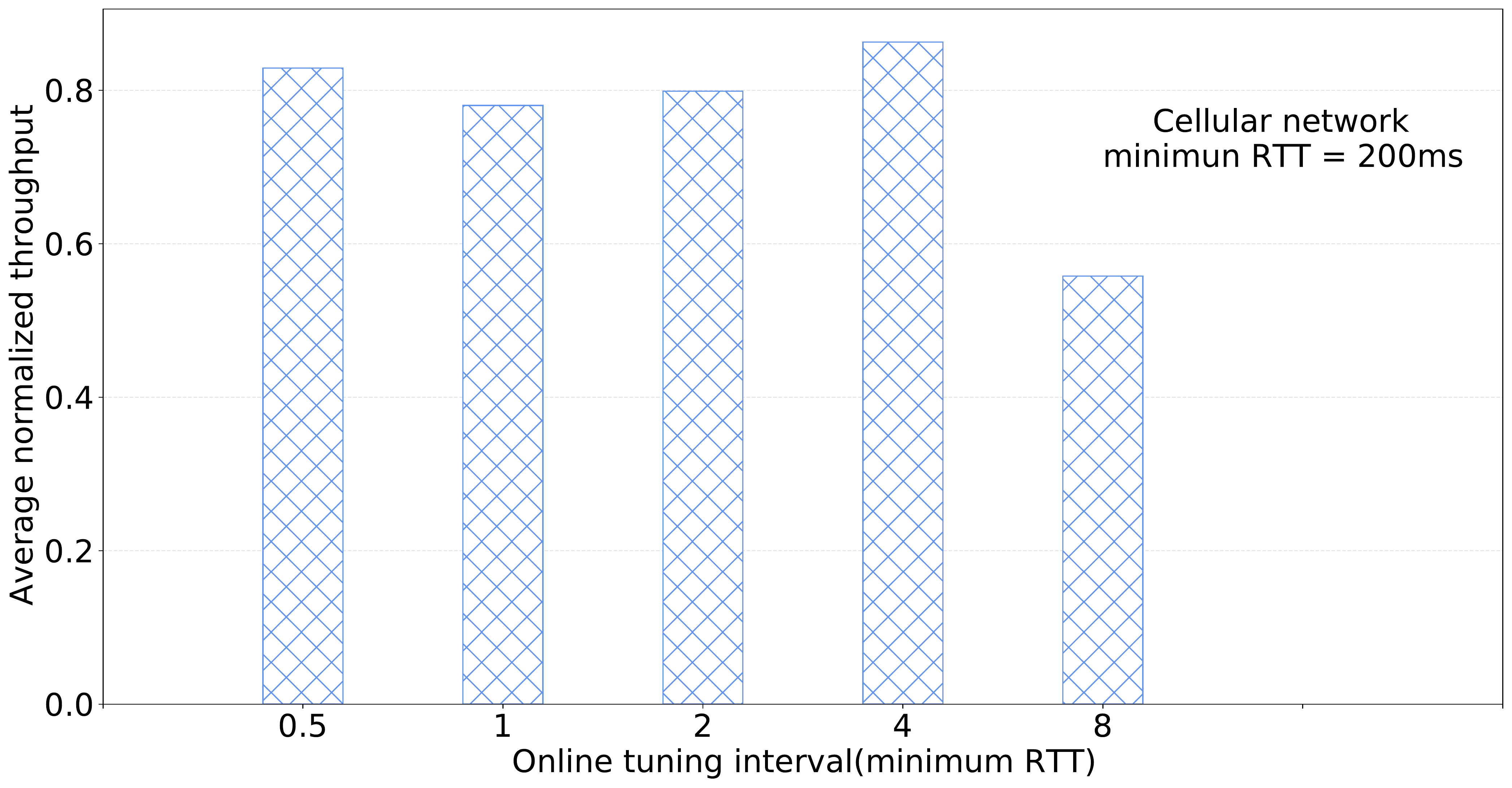}
		\caption{\label{fig:online-decision}The average normalized throughput across different tuning intervals.}
	    \end{minipage}
	   % \vspace{-0.1in}
\end{figure*}

\subsection{Decision interval and overhead}
\label{eval:overhead}
    DeepCC agent is trained on a PC with an Intel Core (TM) i7-6850k 3.6 GHz with 8 GB memory. It requires approximately two million training iterations to obtain the near Pareto optimal policies that perform well enough. In total, the training process took about 15 hours.
    At runtime, it takes about 0.5$\sim$0.9 ms on average to get the \emph{action} (i.e., the model inference time) or update the \emph{goal}.
    We carefully set the decision interval as 1/2 minimum RTT and the online tuning interval as four RTT.

    Here, we investigate the impact of different decision intervals on the performance. Intuitively, the decision interval determines the frequency of model inference. To evaluate the impact on average throughput, we set different intervals across different RTTs environments. Fig.\ref{fig:offline-decision} depicts the results over different decision intervals. There are two takeaways here. First, fixed decision intervals are not advisable. The larger or smaller value of the decision interval can not achieve better performance. Second, the decision interval should change dynamically in different RTT environments. According to the results, we empirically set 1/2 minimum RTT as DeepCC's decision interval.

    % Benefiting from that, the state can reflect the changes in network conditions in time. Meanwhile, it provides sufficient time for the action to be fully executed in the case of no congestion, which means the data sent by the sender can reach the receiver, thereby the estimation of reward is more accurate.
    Online tuning interval determines the {execution frequency} of the gradient descent. If the interval is too large, DeepCC cannot adjust the policy in time according to the network condition to satisfy application requirements. On the other hand, if the interval is too small, {DeepCC will adjust its policy too frequently thus inccuring unstablizaiton}.
    To evaluate the impact of tuning intervals, we set experiments running the online tuning module with different  intervals from half to 8 RTTs.
    % The online goal tuning is triggered by four RTTs rather than every ACK according to our experimental results.
    The results in Fig.\ref{fig:online-decision} show that the average normalized throughput under 4 RTT is the best than that of other intervals. Therefore, we use 4 RTT as online tuning interval in this paper.
    % We omit the results due to space limitations.}.

\begin{figure}[t]
		\centerline{
	    \includegraphics[width=0.99\linewidth]{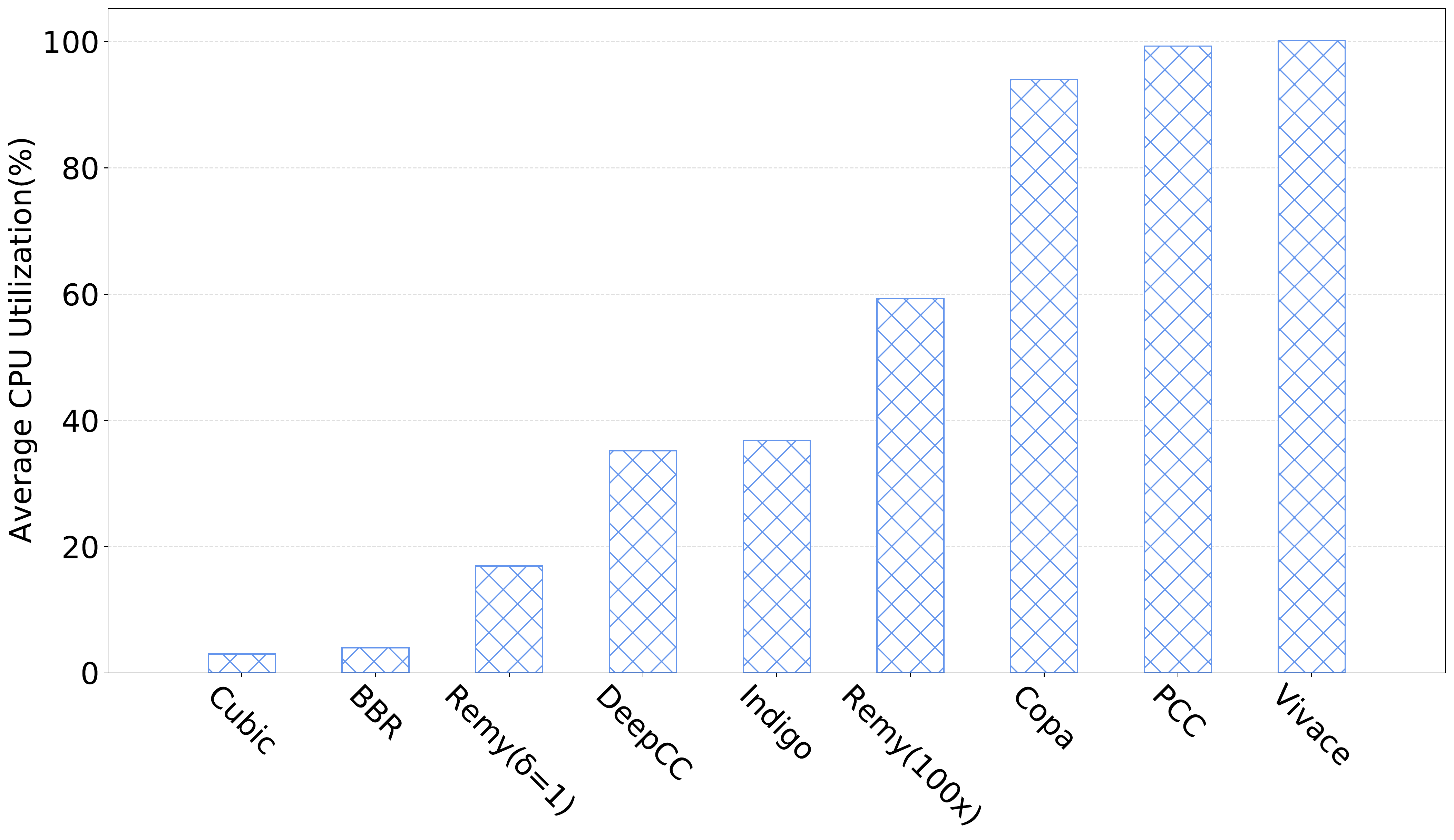}}
		\caption{\label{fig:cpu_using}Overhead of DeepCC compared with other schemes.}
% 		\vspace{-0.15in}
	\end{figure}

    \noindent \textbf{Overhead. }To investigate the overhead of DeepCC and compare it with the off-the-shelf congestion controls, we set up  an emulated network with 12 Mbps bottleneck link and 60 ms RTT for 60 seconds and send traffic from the sender to the receiver. We measure the average CPU utilization of these schemes on the sender. On account of Cubic and BBR implemented in the kernel, we evaluate the CPU utilization of iperf for sending their traffic. Results are shown in Fig.~\ref{fig:cpu_using}. It is worth nothing that the overhead of different offline-trained models is not the same due to their different complexity. For example, the overhead of Remy(100x)\footnote{The model with a 100x range of link rates is provided by \cite{sivaraman2014experimental}.} is 40\% higher than that of Remy($\delta=1$)\footnote{The model which $\delta$ represents the relative importance of latency is provided by \cite{Winstein2013Remy}.}. As we expected, {DeepCC, similar} to Indigo, achieves a lower overhead than that of other schemes except for BBR and Cubic. If we implement DeepCC in a more efficient language (such as C or C++ language) instead of Python, or convert DeepCC agent into a decision tree\cite{Metis} in the Linux kernel, it {could be possible to achieve even lower overhead. We leave this as our future work.}
% \subsection{DeepCC deep dive}
%    It is well-known that the learning-based approaches are black box lacking a theoretical understanding of policies from the agents. To investigate that how DeepCC achieves adaptiveness, we describe the behavior of DeepCC agent under two scenarios which are that the bandwidth is suddenly reduced and the bandwidth is suddenly increased. The results are shonw in Fig.\ref{fig:deepdive}. XXXXX

\begin{figure*}[t]
	\centerline{\includegraphics[width=0.95\linewidth]{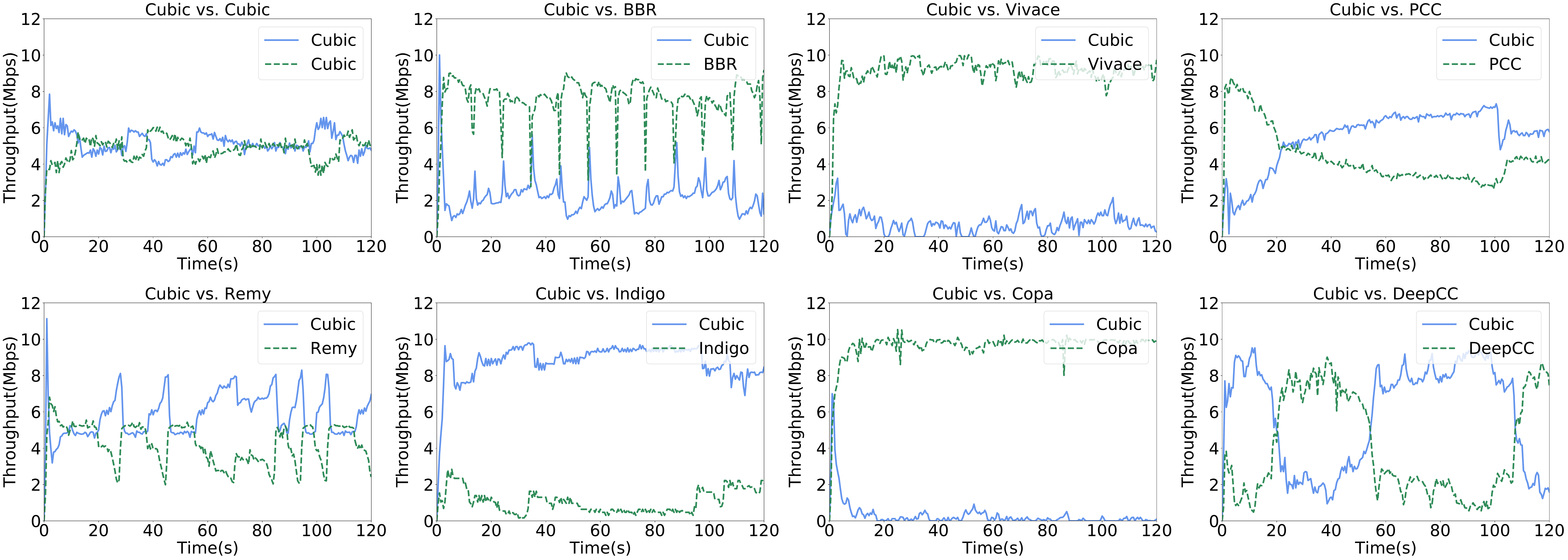}}
	% \vspace{-0.15in}
	\caption{\label{fig:friendliness}Throughput dynamics of different congestion controls competing the bottleneck link with Cubic.}
	\vspace{-0.15in}
\end{figure*}

\subsection{Coexistence of DeepCC and non-DeepCC flows}
	\label{eval:fairness}
	To evaluate DeepCC with competing flows, we set up our testbed to experiment with the coexistence of DeepCC or non-DeepCC flows. We use two hosts as the sender and receiver respectively.
	They are connected through a router running OpenWrt\cite{openwrt}. We use tc\cite{TC} in OpenWrt to regulate the bottleneck link.
	% Bottleneck router buffer size is set to the BDP.
	% In this section, we experimentally show the dynamic behavior among DeepCC and non-DeepCC flows.

	\noindent\textbf{Coexistence with non-DeepCC flows.}
	\label{eval:friendly}
	To evaluate the DeepCC's friendliness, we examine different target flows of DeepCC competing with Cubic flows on our testbed. We choose Cubic flow as the reference flow rests on the fact that Cubic is the default deployment in Linux kernel. In our experiments, we start simultaneously two flows from the sender to the receiver using different schemes including Cubic, BBR, Remy, Indigo, PCC, Vivace, and DeepCC.

	As shown in Fig. \ref{fig:friendliness}, we report the average throughput achieved by each scheme and Cubic throughout time. The results indicate that BBR, Vivace and Copa are aggressive and get nearly all the bandwidth from the Cubic flow. However, when completing with Remy, Indigo and PCC, Cubic is aggressive, while PCC's share of bottleneck link's bandwidth changes from the high to low and does not grow in the presence of Cubic.

	In the first few seconds, Cubic quickly grabs the bandwidth. When the queue is full and packet loss occurs due to congestion, Cubic reduces its cwnds. However, at this time, BBR is not sensitive to packet loss and still takes two times of BDP as its cwnds. Therefore, BBR flow can fully leverage the queue while the Cubic flow can not share fairly the bandwidth due to its {sensitivity} to packets loss.
	PCC, Vivace and Copa set their sending rate based on their predefined utility function. They require good start point and need time to find the good sending rates when competing with the Cubic flow. In other cases, Remy and Indigo are delay-sensitive congestion control schemes. They focus on the delay and have lower rate while Cubic can achieve high bandwidth. In above cases, the friendliness between these schemes {and Cubic} is not desirable.

	In Fig.\ref{fig:friendliness}, Cubic flow can share fairly the bandwidth with Cubic flow. However, DeepCC ({the target is set with} 5 Mbps, 180 ms, 1\%) mainly uses the available bandwidth outside of the cubic flow, and achieves a proportional fairness with the cubic flow. When the sending rate of Cubic decreases, the available bandwidth grows and DeepCC increases the sending rate accordingly to obtain higher bandwidth. When the sending rate of Cubic increases, the available bandwidth reduces and DeepCC also reduces the sending rate. {Although the obtained bandwidth of DeepCC fluctuates, in a long run it achieves similar average throughput with that of Cubic.}
\begin{figure}[t]
	% \vspace{-0.1in}
		\centering
		\includegraphics[width=0.95\linewidth]{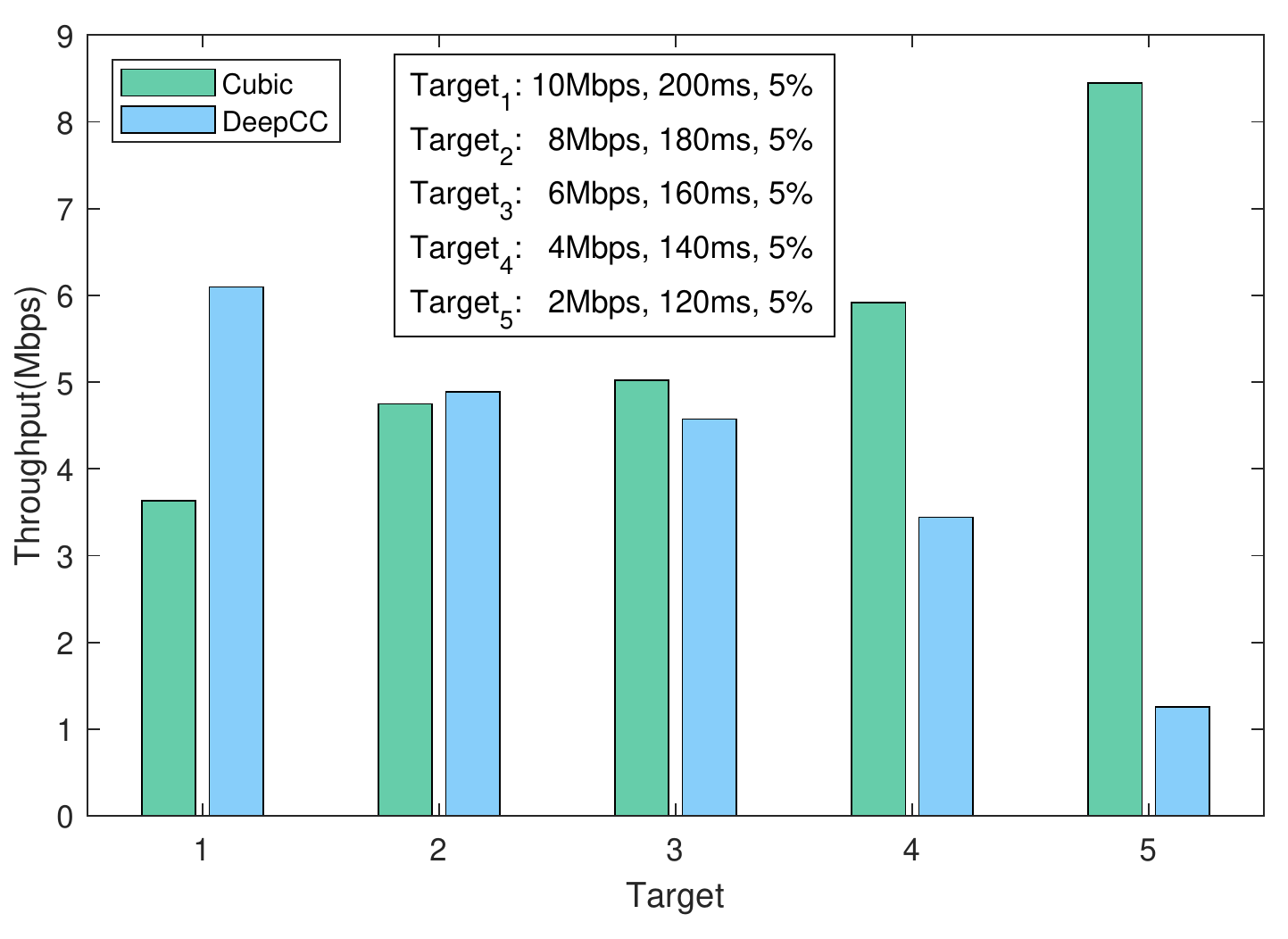}
		% \vspace{-0.15in}
		\caption{\label{fig:friendly}Throughput of different DeepCCs vs. Cubic.}
% 		\vspace{-0.15in}
\end{figure}

	Furthermore, we use {DeepCC with different targets} to investigate {its} completing behaviors. Fig. \ref{fig:friendly} shows that the strong preemption of DeepCC in high-throughput mode benefits from the ability of utilizing the capacity of the bottleneck. It has a higher throughput than that of Cubic. When the target is in the low-latency mode, DeepCC has a weaker preemption of available bandwidth than that of Cubic. When the target is set in the middle, DeepCC achieves a better friendliness {on} average throughput with competing Cubic flows. As expected, {DeepCC with different targets achieves varying proportional} share of the bandwidth.

\begin{figure}[t]
	\centerline{\includegraphics[width=0.95\linewidth]{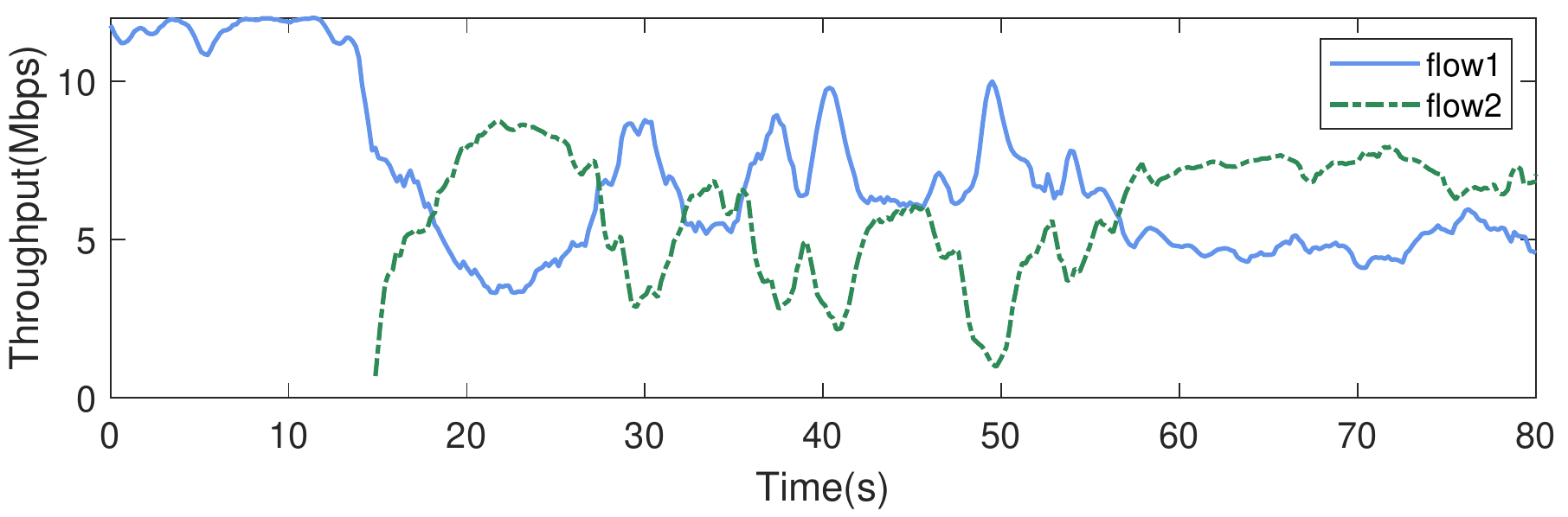}}
	% \vspace{-0.15in}
	\caption{\label{fig:fair-same-tar}Dynamic behavior of {competing} DeepCC flows with the same target.}
	% \label{single-scenario-multi-objective}
	\vspace{-0.15in}
\end{figure}

	\noindent\textbf{Coexistence of DeepCC flows with the same target.}
	To understand the {behavior of DeepCC facing} other DeepCC flows, we set up two competing {DeepCC} flows with the same target, e.g., high-throughput target. The data transmission of the two flows initiates sequentially with a 15s interval and each flow transmits continuously for 80s. Fig. \ref{fig:fair-same-tar} depicts the fairness property between two {DeepCC} flows with the same target. As expected, DeepCC can achieve an acceptable {fairness} with other DeepCC flows. If there are two long flows competing with the bandwidth, DeepCC can achieve a {good} fairness characteristic through a period of adjustment.
% 	We can see that DeepCC flows with the same target converge to a similar throughput via the continuously \emph{goal} tuning.

\subsection{Robustness}
	\label{common properties}
		In the following, we evaluate the robustness and convergence of DeepCC. The emulated network link is set as a steady link with 12 Mbps, 60 ms RTT, and 90 KB of buffer.

	\noindent\textbf{Robustness to packet loss. }
		When facing the lossy network condition, the loss-based and delay-based schemes often perform poorly, which is often shown as a sharp decline in throughput. On the contrary, the learning-based schemes can combat more non-congested packet loss since they do not take packet loss or delay as the explicit congestion signal.

\begin{figure}[t]
        \includegraphics[width=0.99\linewidth]{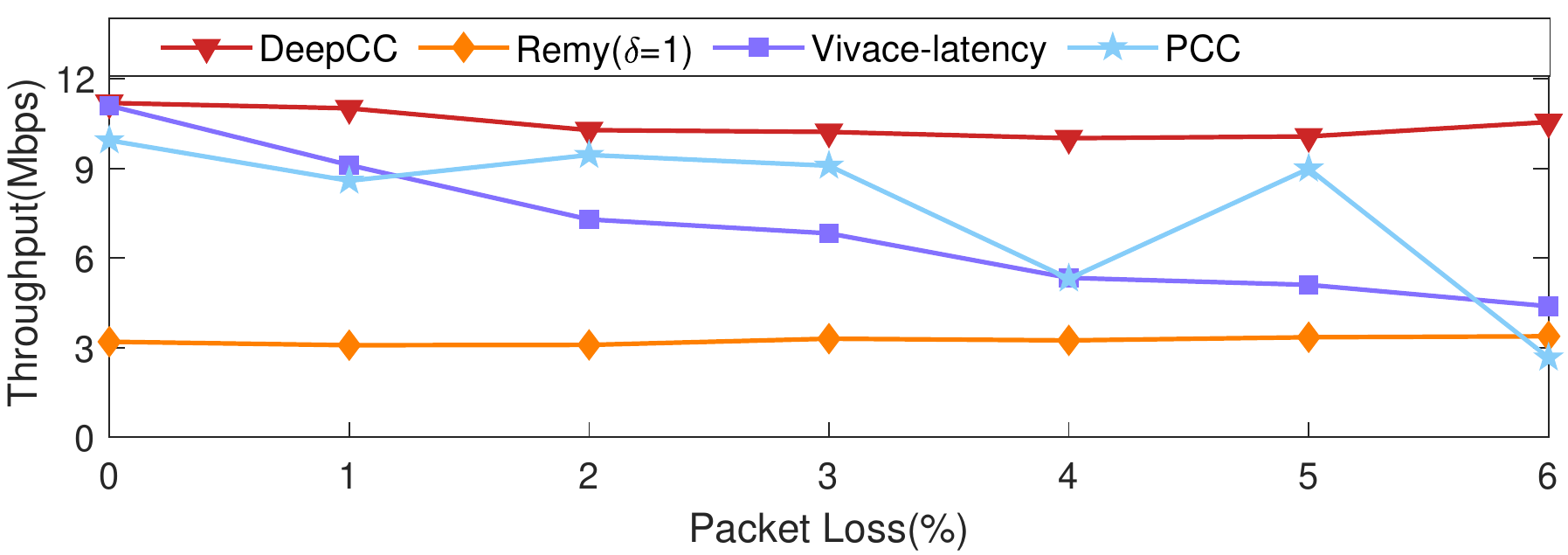}
		\caption{\label{packets-loss}Robustness to stochastic packet loss.}
% 		\vspace{-0.15in}
\end{figure}

		To evaluate the robustness to packet loss of DeepCC, we set up a steady link with a stochastic loss rate ranging from 0\% to 6\%. The target for DeepCC is set as 11 Mbps throughput, 100 ms delay, and 5\% loss rate. The target loss rate represents the maximum tolerable link loss rate.
		As shown in Fig.~\ref{packets-loss}, Remy is insensitive to packets loss, whereas it only achieves low throughput. PCC maintains throughput until the packet loss rate reaches 5\%, i.e., the predefined loss tolerance included in its objective function. The throughput of Vivace gradually decreases as the stochastic loss rate increases.

		DeepCC remains insensitive to random packet loss and obtains consistent high throughput. This is because the goal (more precisely $g^{(3)}$) is not updated by the online tuning algorithm when the observed loss rate is less than that of target. Thus DeepCC ignores the random loss and runs to pursue the other two dimensions of the target (i.e., throughput and delay). If the observed loss rate is higher than that of target and the other two dimensions of measurement are satisfied with the target, DeepCC does not update the gradient according to the constraint of loss rate gradient in the online tuning algorithm.

\begin{figure}[t]
% 		\centerline
        \includegraphics[width=0.99\linewidth]{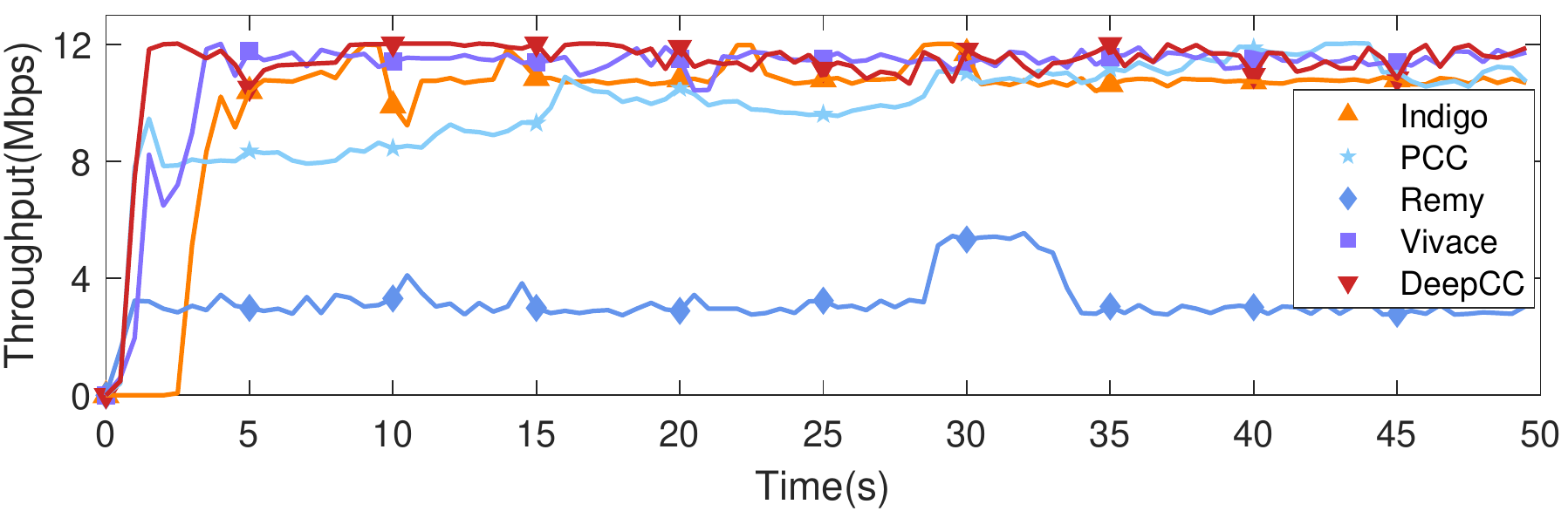}
		\caption{\label{convergence time}Convergence time {of} different learning-based schemes.}
% 		\vspace{-0.15in}
\end{figure}

	\noindent\textbf{Convergence time of learning-based schemes. }
		\label{convergence-time}
		Here the convergence time refers to the time for the learning-based algorithms to reach a stable state.
		Although online schemes could react timely to the variable network conditions, the performance may be greatly impacted by convergence time. To evaluate DeepCC, we set the \emph{target} value {$T$} as 12 Mbps throughput, 100 ms maximum delay, and 1\% loss rate. Fig.~\ref{convergence time} illustrates the convergence process of several congestion control schemes with 0.5s granularity. As the results showed, DeepCC has similar stable convergence behavior as the offline algorithms, i.e., Indigo and Remy (ver. $\delta$=1.0), but the throughput of DeepCC is higher than that of them. Compared with online learning schemes, DeepCC has the same ability to quickly lift throughput to 12 Mbps as Vivace (ver. latency), but the throughput of DeepCC oscillates less than that of it. In contrast, PCC takes the longest time to reach the throughput ceiling. Benefiting from offline and online learning, DeepCC gains a good start point from offline training and quickly adapts to network conditions through online fine-tuning.

\section{Discussion}
\label{discussion}
% \vspace{-0.05in}
    DeepCC is a flexible congestion control generated by multi-objective learning and dynamically tuning policy, i.e., optimization objective for different network conditions and application requirements. Although the performance of DeepCC achieved is encouraging, some issues remain to be solved. 
    
    (1){ Multi-objective representation.} 
    {In this paper, DeepCC adopts the form of weighted objective functions about three metrics, i.e., throughput, delay, and packet loss, as its offline optimization objective. If applications also concern other performance metrics, such as jitter, the objective function of DeepCC can be easily extended to support it. Further, we leave the cases that the optimization objective involves more complex forms or constraints for our future work.}  
    % If applications have more complex and sophisticated objectives or constraints, the multi-objective function should be handling these cases. 
    % In future work, we will consider adding other metrics in the multi-objective function, such as power consumption, buffer size, jitter, etc..

    % \noindent \textbf{\zl{Unapplicable network scenario. }}
    % Although DeepCC tries to generalize for different network conditions, there are some scenarios (e.g., the shallow-buffered network) that the benefits of applying DeepCC are limited. The reason is that DeepCC attempts to satisfy the application requirements by controlling the utilization of the bottleneck buffer. However, the controllable space is narrow if the buffer is too shallow. Under this circumstance, DeepCC would tune to the similar policy even facing different requirements, which degenerates from the multi-objective to the single objective optimization. 
    % % \zl{(See in \S\ref{eval:fairness})}. 

    (2){ Overhead of large-scale deployment. }
    DeepCC can be deployed on the server-side and maintain the client unchanged. The control decision of DeepCC is obtained from the neural networks at the server.  When multiple concurrent connections are established at the server, the memory and CPU consumption will increase, and thus influence the model inference time. The server can mitigate this impact by 1) running the model inference as a service (e.g., tensorflow serving\cite{olston2017tensorflow}) with multiple instances, 2) performing the load balancing, and 3) limiting the maximum number of concurrent connections.

    % \noindent \textbf{Practical deployment and overhead.} DeepCC can be deployed on the server side.
    % The control decision of DeepCC is obtained from the neural networks at the server side while the client remains unchanged. 
    % % Therefore DeepCC can easily support a broad range of client devices without modifications. 
    % Additionally, the neural networks introduce complexity and overhead at the endpoint, but our results suggest that the additional latency could have negligible impact on the performance by setting a proper decision interval (\S\ref{evaluation-setup}). Further, the overhead could be reduced using compressed neural networks\cite{han2015deep}. 
    % We also leave research questions regarding the friendliness between DeepCC and legacy CC algorithms and the production system to the future. 

     (3){ Coexistence of DeepCC flows with different targets.}
    A potential concern for DeepCC is how high-throughput flows affect the low-latency flows. 
    We experiment with our testbed using a bottleneck link of 10 KB of buffer \cite{RouterBuffer}. 
	The results show that the performance of these flows achieves a similar throughput and delay respectively. Specifically, the queueing delay is close to zero. Due to the shallow-buffered network, the controllable space of DeepCC is narrow. Under this circumstance, DeepCC would tune to a similar policy even facing different targets, which degenerates from the multi-objective to the single-objective optimization.
	However, these results can not fully reflect the performance of DeepCC in the real networks since the last mile is typically the speed bottleneck in communication networks\cite{lastMile}. Therefore, DeepCC can benefit from different bottlenecks to satisfy different targets in \S\ref{eval:wild}. 
\section{Related Work}
    Congestion control has been continuously studied since the advent of computer networks. As conventional schemes, such as Cubic~\cite{XU2005CUBIC} and Vegas~\cite{OMalley1994Vegas}, were designed for general purposes with a ``best effort'' mentality, they show disadvantages in satisfying modern applications that pose strict requirements on networking. Despite some efforts in~\cite{Leong2017PropRate} are made to realize explicit control for buffer delay, it only works well in the cellular network and cannot extend to other network scenarios or metrics e.g., throughput or loss rate.
    % Traditional congestion control schemes (e.g., Cubic~\cite{XU2005CUBIC}, Vegas~\cite{OMalley1994Vegas}), inherently cannot satisfy various application requirements since they lack application awareness and can only work in a limited operating range. Recently, a buffer-based algorithm~\cite{Leong2017PropRate} attempts to regulate the utilization of the bottleneck link by setting the target buffer delay. Although it supports to optimize for an explicit buffer delay, it only works well for cellular network and cannot satisfy other application requirements (e.g., throughput or loss rate).
    
    Encouraged by the successful experience of machine learning in other fields\cite{Silver2016Mastering,Mao2016Resource,Mao2017Pensieve}, network researchers turn to learning-based approaches.
    Rather than leveraging the control rules designed by humans, they propose to use objective-based approaches to guide the decision-making process in a network environment. Many solutions are raised in this way. 
    For example, Remy~\cite{Winstein2013Remy} and Indigo~\cite{Francis2018Pantheon} perform optimization by learning CC rules {offline}. Once being trained, neither Remy nor Indigo can be adjusted to satisfy the requirements without retraining.
    PCC~\cite{Dong2015PCC} and Vivace~\cite{Dong2018Vivace} depend on {online} learning to make right decisions. Owing to online learning, PCC and Vivace are able to provide no-regret guarantees whereas they do not customize and adapt to the requirements. Orca\cite{Orca} designs two levels of control which combines Cubic and a agent. The agent learns the factor of cwnds for Cubic based on DRL algorithm. Though Orca can achieve high performance in some scenarios, it can not guarantee the application-specified demands.
\section{Conclusion}
% We proposed DeepCC, a congestion control framework based on offline and online learning. 
% Our work bridges this gap between application performance requirements and congestion control and builds on these multi-objective learning optimization insights to develop the DeepCC framework. 
% DeepCC leverages a novel multi-objective optimization DRL algorithm to offline learn various rate control policy and automatically achieves desired outcomes with the multi-dimensional gradient-ascent-based online learning method. Our approach not only works well for different network scenarios but also has the benefit of automatically tuning application performance on a wide range of performance trade-off points according to application requirements. 
To bridge the gap between application requirements and congestion control, we propose DeepCC, a congestion control that combines the ideas from both offline learning and online tuning. DeepCC leverages a novel multi-objective DRL to learn the multi-objective control policy offline and automatically achieves desired outcomes with the gradient-based online tuning method. The experiment results show that our approach not only achieves a wide range of performance trade-offs but also works well for untrained network scenarios.

% or


% \appendices

% % you can choose not to have a title for an appendix
% % if you want by leaving the argument blank
% \section{}

\ifCLASSOPTIONcompsoc
  \section*{Acknowledgments}
\else
  \section*{Acknowledgment}
\fi

The authors would like to thank MingLong Dai and Jing Liu from Beijing University of Posts and Telecommunications for their valuable suggestions and great efforts on the early version of this work.

\ifCLASSOPTIONcaptionsoff
  \newpage
\fi



%\bibliographystyle{IEEEtran}


% \end{thebibliography}
\bibliographystyle{IEEEtran}
\bibliography{DeepCC-main}
\end{document}